\documentclass[useAMS,fleqn,usenatbib,a4paper]{mn2e} 

\pdfoutput=1
\usepackage{times} 
\usepackage{graphicx}
\usepackage{amsmath}
\usepackage{amssymb}
\usepackage{fixltx2e}

\renewcommand{\vec}[1]{\ensuremath{\mathbf{#1}}} 
 
\makeatletter
\providecommand*{\diff}{\@ifnextchar^{\DIfF}{\DIfF^{}}}
\def\DIfF^#1{\mathop{\mathrm{\mathstrut d}}\nolimits^{#1}\gobblespace}
\def\gobblespace{\futurelet\diffarg\opspace}
\def\opspace{\let\DiffSpace\!\ifx\diffarg(\let\DiffSpace\relax
                                                  \else\ifx\diffarg[\let\DiffSpace\relax
                                                  \else\ifx\diffarg\{\let\DiffSpace\relax\fi\fi\fi\DiffSpace}
\newcommand{\deriv}[2]{\frac{\diff #1}{\diff #2}}

\newcommand{\grad}[1]{\ensuremath{{\nabla} #1}}

\providecommand*{\unit}[1]{\ensuremath{\mathrm{\,#1}}}

\newcommand{\boldtxt}[1]{#1}

\title[Gas and dust in SPH]{Smoothed particle hydrodynamics simulations of gas and dust mixtures}

\author[Booth, Sijacki \& Clarke]{R.~Booth,$^{1}$\thanks{E-mail: rab200@ast.cam.ac.uk} D.~Sijacki,$^{1,2}$ \& C.~Clarke$^{1}$\\
$^{1}$Institute of Astronomy, University of Cambridge, Madingley Road, Cambridge, CB3 0HA, United Kingdom\\
$^{2}$Kavli Institute for Cosmology Cambridge, University of Cambridge, Madingley Road, Cambridge, CB3 0HA, ​United Kingdom}

\pagerange{\pageref{firstpage}--\pageref{lastpage}} \pubyear{2015}

\begin{document}

\label{firstpage}

\maketitle

\begin{abstract}
\boldtxt{We present a `two-fluid' implementation of dust in smoothed particle hydrodynamics (SPH) in the test particle limit. The scheme is able to handle both short and long stopping times and reproduces the short friction time limit, which is not properly handled in other implementations. We apply novel tests to verify its accuracy and limitations, including multi-dimensional tests that have not been previously applied to the drag-coupled dust problem and which are particularly relevant to self-gravitating protoplanetary discs. Our tests demonstrate several key requirements for accurate simulations of gas-dust mixtures. Firstly, in standard SPH particle jitter can degrade the dust solution, even when the gas density is well reproduced. The use of integral gradients, a Wendland kernel and a large number of neighbours can control this, albeit at a greater computational cost. Secondly, when it is necessary to limit the artificial viscosity we recommend using the \citet{Cullen2010} switch, since the alternative, using $\alpha \sim 0.1$, can generate a large velocity noise up to $\sigma_v \lesssim 0.3 c_s$ in the dust particles. Thirdly, we find that an accurate dust density estimate requires $>400$ neighbours, since, unlike the gas, the dust particles do not feel regularization forces. This density noise applies to all particle-based two-fluid implementations of dust, irrespective of the hydro solver and could lead to numerically induced fragmentation. Although our tests show accurate dusty gas simulations are possible, care must be taken to minimize the contribution from numerical noise.}
\end{abstract}


\section{Introduction}
The growth of planets out of km-sized planetesimals remains an attractive 
scenario for both terrestrial planet formation and, in the core accretion
model, also as a necessary step in the creation of gas giant planets.
Nevertheless -   regardless of  whether these planetesimals form by direct gravitational collapse or agglomerative growth -  
there are a number of problems associated with the formation and
retention of planetesimals in protoplanetary discs (for a recent review of the subject see \citet{Johansen2014}).

For example, although there is observational evidence at mm and cm wavelengths
for rather rapid
grain growth \citep{Wilner2005,Rodmann2006,Ricci2010},  agglomerative growth 
 beyond $\sim $ cm-sizes is inefficient,  since collisions tend to result in fragmentation or bouncing rather than sticking and grain growth. Particles in this
size range are said to be `critically coupled' to the disc gas, having
values of the Stokes number ({\rm St}: the ratio of the timescale for gas drag to the
dynamical time) of around unity. Drag causes such particles to 
lose angular momentum to
the gas (which orbits at slightly sub-Keplerian velocity on account of  being  
 partially  supported  by outwardly directed pressure gradients) and to
 drift towards the star on time-scales of 100 to 1000 orbits \citep{Weidenschilling1977}. Once the drifting particles reach the snow-line they will sublimate, enhancing the local density in vapour form \citep{Cuzzi2004}, removing most of the solids from the outer disc within $1{\rm \,Myr}$ \citep{Brauer2008}.  

The loss of solids by radial drift  can potentially be overcome if pressure does not increase monotonically with decreasing radius in the disc  and if instead some process can create and sustain local  pressure maxima.  In this case solid material can be trapped at the maximum (since regions of outwardly increasing pressure drive outward migration of solids by the same argument).  For example, the pressure maxima  at the outer edge of gaps formed by massive planets have been proposed as an effective dust trap  \citep{Lyra2009}, and this  has been invoked to explain the asymmetric dust around Oph IRS 48 \citep{vanderMarel2013}. During the early phase of protoplanetary disc evolution, while the disc is still massive enough that gas self-gravity is important,  spiral structures may likewise slow or halt the radial drift of dust grains, concentrating dust in  the pressure maxima associated with spiral arms \citep{Rice2004,Rice2006,Gibbons2012}. If this focusing  of dust into the spiral arms can raise its density by two orders of magnitude then it becomes the dominant mass component locally and  direct gravitational collapse in the dust layer - creating km scale planetesimals - may ensue.

If planetesimals do form in self-gravitating discs, then the spiral structure will drive eccentricities to $e \gtrsim 0.1$. The high velocity dispersion (and hence suppression of gravitational focussing) renders collisions negligible and hence planetesimals should survive the self-gravitating phase \citep{Walmswell2013}. Planetesimal formation is most likely to occur at radii greater than $10\,{\rm AU}$ , where the relatively short cooling time is linked to a larger amplitude of spiral structure \citep{Clarke2009}.  The direct conversion of grains that are critically coupled (with size  1 -- 10\unit{cm} in this region of a self-gravitating disc) into planetesimals offers the attractive prospect of by-passing the problems of agglomerative growth over intermediate size scales.

 Since the mechanism requires a significant fraction of the dust to be in large grains that have a Stokes number, ${\rm St} \approx 1$, the success of this model depends on their survival during the self-gravitating phase. Their survival is sensitive to relative velocity of collisions between particles, since collisions with velocities greater than 1 -- 10${\rm \,m\,s}^{-1}$ result in fragmentation \boldtxt{\citep{Guttler2010}}. Since \citet{Rice2004} found velocity dispersions of order the sound speed, which is approximately $500 {\rm \,m\,s}^{-1}$ at $30{\rm \,AU}$, collisions may destroy the grains before a sufficient density has been built up for collapse to occur. 

The results of competing grain growth and fragmentation processes require detailed knowledge of the velocity distributions of the dust grains during the self-gravitating phase, and this  remains  uncertain. \boldtxt{In particular, early simulations of self-gravitating discs were studied using Smoothed Particle Hydrodynamics (SPH), which had problems related to the need to reduce artificial viscosity in order to avoid a situation where the heating by the gravitational instability is exceeded by the action of artificial viscosity on Keplerian shear \citep{Lodato2004, Rice2004, Rice2006}. The amount of artificial viscosity in SPH is  typically controlled by two parameters, $\alpha$ and $\beta$, which govern artificial viscosity terms that are
linear and  quadratic in the relative velocity, respectively\footnote{For a detailed description of artificial viscosity used in SPH, see \citet{Price2012}; although note that the code used in this work (\textsc{gadget-2}) uses a slightly different form \citep{Springel2005}.}. To reduce the viscosity often $\alpha$ was reduced by an order of magnitude from the typical values of $\alpha = 1$ and sometimes $\beta$ was also reduced \citep{Lodato2004, Rice2004, Rice2006, Meru2011, Meru2012}.} Subsequently it has been shown that such a choice  fails to generate enough entropy in shocks and results in considerable noise in the gas velocities (\citealt{Meru2011, Meru2012, Rice2014}; see also \citealt{Lattanzio1986,Lombardi1999}  and \citet{Thacker2000} who suggest $\alpha \gtrsim 0.7$, even for weak shocks). The velocity noise is likely to be transmitted to the dust component via drag forces and this means it is not entirely clear that the velocity dispersion measured in a simulation is physical. Recent shearing-box simulations go some way towards  addressing the issue \citep{Gibbons2012, Gibbons2014} and  also find a  velocity dispersion of dust particles over the entire box of $v \sim c_s$. However, the collision velocities are sensitive to the correlation structure in the velocity distribution, a quantity that has not been derived from these simulations.

With recent developments in computational techniques, accurate simulations of 
the dynamics of gas-dust mixtures are within reach \citep{Johansen2007, Laibe2012a,Laibe2014a,Loren-Aguilar2014}. However, while there have  been significant efforts to verify codes involving the  simulations of turbulence and the streaming instability \citep[e.g.][]{Bai2010a}, there have been few ways to verify that  codes capture the velocity distribution of solids in the context of self-gravitating discs.

In this work we conduct a series of tests that can be used to assess code performance in this respect. While some of these (e.g. the \textsc{dustybox} and \textsc{dustyshock} tests described in \citet{Laibe2011}) have previously been used in this way, we also introduce two tests that have not previously been examined with respect to the modelling of coupled dust. These two tests involve shocks in 2D: the implosion test of \citet{Hui1999} and the modelling of a rigidly rotating spiral mode in an isothermal gas disc \citep{Roberts1969,Shu1973}. The latter in particular was chosen because it bears some resemblance to the problem of interest (since it involves centrifugally supported gas interacting with a pattern of spiral shocks) but - unlike the {\it self-}gravitating disc problem - is amenable to
an analytic solution for the gas that is steady in the co-rotating frame. For each test we quantify the velocity structure in the dust and how this is affected by a range of numerical choices. Finally, we summarise our conclusions in terms of a set of recommendations and requirements for the simulation of dust in self-gravitating protoplanetary discs. 

\section{Gas - Dust dynamics}

The dynamics of solids in protoplanetary discs is governed by the interplay of drag forces with Keplerian motion. The force on a solid particle of mass $m_d$ is given by
\begin{equation}
\vec{F}_d = m_d \frac{\rm d \vec{v}_d}{\rm d t} = -  K_s (\vec{v}_d - \vec{v}_g) + m_d \vec{a}_{d,{\rm ext}}, \label{Eqn:DragLaw}
\end{equation}
and the corresponding equation for the gas is
\begin{equation}
\frac{{\rm d} \vec{v}_g}{{\rm d} t} =  - \frac{\nabla P}{\rho_g} + \frac{\rho_d}{m_d \rho_g} K_s (\vec{v}_d - \vec{v}_g) + \vec{a}_{g,{\rm ext}}, \label{Eqn:GasLaw}
\end{equation}
where $\vec{v}_g$ and $\vec{v}_d$ are the local gas dust and gas velocities, $\rho_g$, and $\rho_d$ are their respective densities, $P$ is the gas pressure and $\vec{a}_{g,{\rm ext}}$ and $\vec{a}_{d,{\rm ext}}$ are additional accelerations. These have been made distinct for the gas and dust to allow for forces that affect both phases, such as gravity and those that only affect a single phase, like viscosity. In these equations $\tfrac{{\rm d}}{{\rm d} t}$ refers to the full Lagrangian derivative in the frame moving with the gas or dust.

The drag coefficient, $K_s$, depends on the Reynolds number, gas density and temperature, which can vary significantly over the different gas and grain properties found in astrophysical problems. \boldtxt{The stopping time, $t_s$, can be defined as the time over which the velocity difference $\Delta \vec{v} = \vec{v}_d - \vec{v}_g$ decays, $\deriv{\Delta \vec{v}}{t} = - \Delta \vec{v} / t_s$, giving
\begin{equation}
t_s = \frac{m_d \rho_g}{K_s (\rho_g + \rho_d)}.
\end{equation}
For negligible dust densities, $\rho_d / \rho_g \ll 1$, this is equal to the stopping time for a single dust particle, $m_d / K_s$.} In protoplanetary discs an approximation for the drag coefficient by \citet{Whipple1972} is generally used. For small grains, where the grain size, $a$, is less than the mean free path, $\lambda$, the drag force can be calculated using the Epstein approximation 
\begin{equation}
K_s = \frac{4 \pi}{3} \rho_g a^2 v_s,
\end{equation}
where $v_s = \sqrt{8 k_B T / \pi \mu}$, $T$ is the gas temperature, $k_B$ is the Boltzmann constant and $\mu$ is the mean-molecular weight. For $a > 9 \lambda / 4$, the drag force can be calculated in the Stokes regime,
\begin{equation}
K_s = \frac{1}{2} C_D \pi a^2 \rho_g | \vec{v}_d - \vec{v}_g |. 
\end{equation}
The coefficient $C_D$ can be approximated by
\begin{equation}
C_D =  \left\{ \begin{array}{l l}
			24 / {\rm Re},       & \quad {\rm Re} < 1, \\
			24 / {\rm Re}^{0.6}, & \quad  1 < {\rm Re} < 800, \\
			0.44,                & \quad {\rm Re} > 800,
			\end{array} \right.
			\label{Eqn:C_D}
\end{equation}
where, ${\rm Re} = 2 a \rho_g | \vec{v} - \vec{v}_g | / \nu$ is the Reynold's number of the fluid and $\nu$ is the molecular viscosity of the gas. Typically in astrophysical applications $\rho_d / \rho_g \ll 1$. In these circumstances the effect of drag forces on the gas can be neglected and the dust particles can be considered as test particles, with $\rho_d = 0$. While there are interesting astrophysical problems in which non-negligible dust density is required, such as the streaming instability \citep{Johansen2007}, the test-particle limit has the advantage of allowing comparison between the simulations and analytical solutions. Furthermore, it allows a more direct comparison with work on self-gravitating protoplanetary discs, in which the test particle limit was used \citep[c.f.][]{Rice2004,Rice2006,Gibbons2012}.

\subsection{Implementation}

A number of different approaches to including the mutual dynamics of gas and solids in SPH have been proposed, which can be broadly separated into single-fluid or two-fluid approaches \citep{Rice2004, Laibe2012a, Laibe2014a,Loren-Aguilar2014}.  In two-fluid approaches the gas and solids are integrated by including separate particle types for the gas and dust grains, while the single-fluid approach considers the dust fraction $\epsilon = \rho_d/ (\rho_g + \rho_d)$ and relative velocity $\Delta\vec{v} = \vec{v_d} - \vec{v_g}$ to be properties of the SPH particles and solves equations for these quantities on an unstructured mesh made up of the SPH particles. 

In the case of high dust to gas ratios, where momentum feedback on the gas from the dust is important \citet{Laibe2014a} have shown that it is essential that all length scales in the problem are resolved for  both the gas and dust
treatments, which places a requirement on traditional two-fluid approaches that the smoothing length, $h \lesssim t_s c_s$, where $c_s$ is the sound-speed. Single fluid approaches appear to avoid this difficulty, but they do so at the cost of the full phase-space information, since only one velocity each of the gas and dust components can be known at each point in space. Conversely, the two-fluid approach is able to capture full phase space information. In the test particle limit the motions of the dust particles are independent of each other and thus their  motions can be calculated accurately as long as the relevant scales in the gas are resolved. Thus the criterion $h \lesssim t_s c_s$ does not apply unless the gas varies on these scales. \citet{Loren-Aguilar2014} have shown this idea holds in general for low dust-to-gas ratios.

Since the full phase-space is essential for evaluating the collisions between grains and the growth of planetesimals, we take a two-fluid approach, which we implement into the SPH code \textsc{gadget-2} \citep{Springel2005}. The two-fluid method has been implemented either by considering pair-wise forces between gas and dust particles \citep{Monaghan1995,Laibe2012a,Loren-Aguilar2014}, or by interpolating the gas properties to location of the dust \citep{Rice2004,Rice2006}. When back-reaction is included pair-wise forces are preferable, since they explicitly conserve angular momentum. However, for the test-particle limit we find a similar performance for both methods (see Appendix \ref{Sec:AppendImpl}). We take the approach of interpolating the gas properties to the location of the dust particles, since it is less sensitive to the underlying gas particle distribution.

The gas properties are interpolated to the location of each dust particle using a kernel sum. For a property $A$, where each gas particle has the value $A_i$, its value at the location of a dust particle $A_D$ can be evaluated via
\begin{equation}
A_D = \sum_j A_j \frac{m_j}{\rho_j} W(|\vec{r}_j - \vec{r}_D|, h_D),
\end{equation}
where $\vec{r}_j$, $m_j$, $\rho_j$ and $A_j$ are the position, mass and density and quantity to be interpolated evaluated at the location of the $j$th gas particle, and $W$ is the smoothing kernel. The sum is over the gas particles neighbouring the dust particles and smoothing length, $h_D$, is set in the same way as for the gas particles, by solving
\begin{equation}
N_\mathrm{NGB} = \sigma h^\nu \sum_j W(|\vec{r}_j - \vec{r}_D|, h_D),
\end{equation}
for $h_D$ at the location of the dust particle $\vec{r}_D$.
\boldtxt{The factor $\sigma h^\nu$ is the volume of the kernel, in three dimensions $\nu = 3$ and $\sigma = 4 \pi /3$; and  $N_\mathrm{NGB}$ is the desired number of neighbours.} The smoothing length and interpolated gas properties are evaluated for all active dust particles after the gas density has been calculated. As a default choice for the kernel, $W(r, h)$, we use the standard cubic-spline kernel and choose $N_\mathrm{NGB} = 50$. In some tests we also consider Wendland kernels. We note this is contrary to recommendation of \citet{Laibe2012a}, but as shown by the tests in the following section, this choice generates the expected forces.

\subsection{Time Integration}
\label{Sec:TimeInteg}
Time integration for gas particles in \textsc{gadget-2} uses a Kick-Drift-Kick leapfrog algorithm, in which the gas velocities and positions are updated alternatively. The positions, $\vec{x}^n$, and velocities, $\vec{v}^n$ at time $t$ are advanced to time $t + \delta t$ via  
\begin{align}
\vec{v}'^n    &= \vec{v}^n  + \frac{\delta t}{2} \vec{a}^n,      \\
\vec{x}^{n+1} &= \vec{x}^n  +       \delta t     \vec{v}'^{n},    \\
\vec{v}^{n+1} &= \vec{v}'^n + \frac{\delta t}{2} \vec{a}^{n+1},
\end{align}
where the \boldtxt{total} accelerations $\vec{a}^n$ are evaluated at positions $\vec{x}^{n}$. The leap-frog scheme is a symplectic, time-reversible integrator for Hamiltonian mechanics, which has the advantage that for closed orbits the energy error can neither grow or decrease, resulting in excellent long term stability even when the truncation error of a single step is relatively high.

For dust particles with ${\rm St} \gg 1$, which follow weakly perturbed Keplerian orbits, the leap-frog integrator is a natural choice. However, for ${\rm St} \ll 1$ explicit time integration schemes require very small time steps. Since in these cases $t_s$ is much shorter than any other time-scale in the problem, this suggests the use of implicit, or semi-implicit approaches that allow $\Delta t \gg t_s$. \citet{Loren-Aguilar2014} extend this idea by using the analytical solution in an operator-split approach. They suggest initially updating the velocities according to non-drag forces and subsequently applying the drag.  The first step is 
\begin{align*}
\vec{v}_d^* &= \vec{v}_d + \vec{a}_d \Delta t, \\
\vec{v}_g^* &= \vec{v}_g + \vec{a}_g \Delta t - \frac{\nabla  P}{\rho} \Delta t,
\end{align*}
\boldtxt{where from here we drop the explicit $\mathrm{ext}$ on $\vec{a}_{d,\mathrm{ext}}$ and $\vec{a}_{g,\mathrm{ext}}$.}
The drag forces are applied making use of the analytical solution for relative velocity,
$ \Delta \vec{v}$, which can be derived in the absence of external forces. For linear drag laws this gives
\begin{equation}
\Delta \vec{v}(t + \Delta t) = \Delta \vec{v}(t) \exp(-\Delta t/ t_s).
\end{equation}
Writing $\xi(t) = 1 - \exp(- t / t_s)$, this generates the final dust velocity via
\begin{equation}
\vec{v}_d(t+\Delta t) = \vec{v}_d^* - \xi(\Delta t)(\vec{v}_d^* - \vec{v}_g^*),
\end{equation}
which is valid in the test particle limit. $\Delta t \gg t_s$ leads to ${\vec{v}_d(t + \Delta t) \rightarrow \vec{v}_g^*}$. This property means that although the scheme is stable for all time-steps, it does not produce the correct terminal velocity unless $\Delta t \ll t_s$. \boldtxt{To see this consider a static atmosphere in which the external forces include only constant gravity, $\vec{a}_{\rm g} = \vec{a}_{\rm d} = \vec{g}$. For a static atmosphere both gas velocity and acceleration must be zero, with pressure balancing gravity, $\nabla P / \rho = \vec{g}$.} For large time-steps $\xi \rightarrow 1$ and $\vec{v}_d(t+\Delta t) \rightarrow 0$, which is not the analytical solution, $\vec{v}_d \rightarrow t_s \nabla P / \rho$.

Instead, we approximate $\vec{v}_g$ as constant in both time and space during the time-step, giving rise to the dust kick
\begin{equation}
\vec{v}_d(t+\Delta t) = \vec{v}_d \exp(-\Delta t/t_s) + (\vec{a}_d t_s + \vec{v}_g)\xi(\Delta t), \label{Eqn:ExactLin}
\end{equation}
which is exact for static atmospheres \boldtxt{and uniform flow} for all $\Delta t / t_s$ and reproduces the short friction time limit. This approximation is as good as the approximation that $\vec{v}_g $ and $\vec{a_d}$ are constant. Only when both the pressure gradients and external forces are zero do the analytical solution and operator split time-steps produce equivalent kicks.

In some cases it may be beneficial to take into account the effect of
motion of the gas on the update for $\Delta \vec{v}$ \boldtxt{(since  uniform gas motion is already taken into account in equation \ref{Eqn:ExactLin})}. The full equation of motion for $\Delta \vec{v}$ in a frame moving with the dust is
\begin{equation}
\frac{{\rm d} \Delta \vec{v}}{{\rm d} t} = - \frac{\Delta \vec{v}}{t_s} + (\vec{a}_d - \vec{a}_g + \frac{\nabla P}{\rho_g}) + \Delta \vec{v} \cdot \nabla \vec{v}_g,
\end{equation}
where the last term takes into account the relative advection between the dust and gas (see equation 27 of \citet{Laibe2012a}). While the advection could be calculated\footnote{
While there may be situations in which the relative advection term is the dominant source of error in equation \ref{Eqn:ExactLin}, there are many situations in which equation \ref{Eqn:DragImplicit} represents an improvement. In the cases where the relative advection term is important, it will be adequately controlled by the Courant-like condition because the velocity should not vary significantly on scales of order of the smoothing length in resolved flow. The exception to this is at shocks, where all schemes are limited to first order anyway.},
 neglecting it and approximating the forces as constant during a time-step  gives
\begin{align}
\vec{v}_d(t+\Delta t) &= \vec{v}_d \exp(-\Delta t/t_s) + \vec{v}_g\xi(\Delta t) + (\vec{a}_g - \frac{\nabla P}{\rho}) \Delta t \nonumber \\
	 &\qquad + \left(\vec{a}_d - \vec{a}_g + \frac{\nabla P}{\rho_g} \right) t_s \xi(\Delta t). \label{Eqn:DragImplicit}
\end{align}
The utility of this approximation can be seen by looking at the limits $t_s \rightarrow 0$ and $t_s \rightarrow \infty$. In the first limit, $\xi \rightarrow 1$ and the dust velocity reduces to the short friction time limit, but using the updated gas velocity. For the $t_s = 0$, the dust velocity becomes exactly the updated gas velocity. In the second limit, $\xi \rightarrow \Delta t / t_s$ and the dust kick reduces to the explicit update for the dust. \boldtxt{Since the approximation of
constant $\vec{a}_d$, $\vec{a}_g$ and $\grad{P}/\rho$ better reflects assumptions made for the integration of the equations of motion for the gas, we recommend the
use of equation \ref{Eqn:DragImplicit}. However, in practice we have used  equation \ref{Eqn:ExactLin} in the results presented. Our tests have shown that the difference in the results presented is small, because in general the SPH forces and interpolation dominate the error.}

\begin{figure}
\centering
\includegraphics[width=0.47\textwidth]{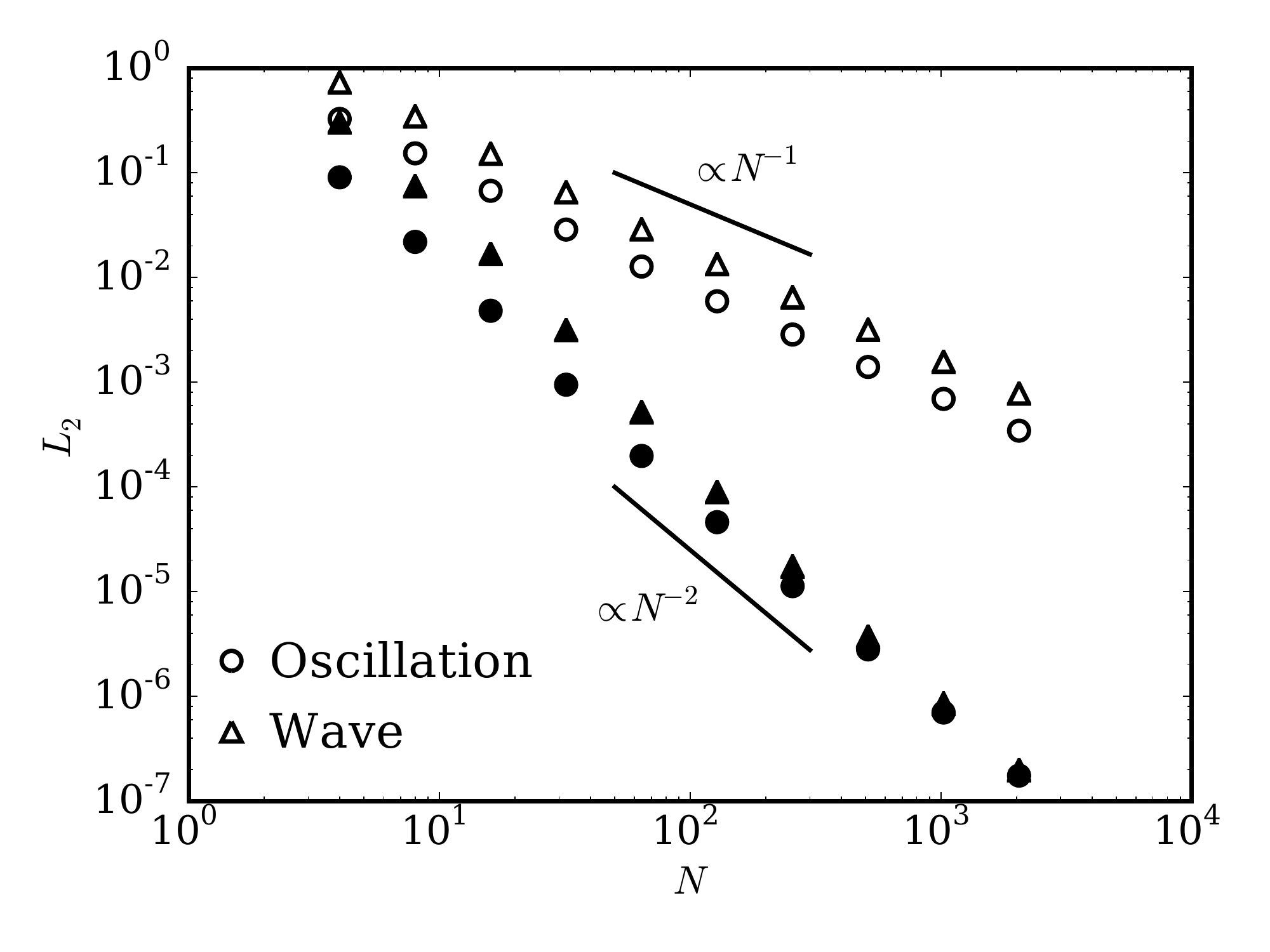}
\caption{Convergence of the drift schemes with the number of steps per period, $N$, in the forced oscillation (circles) and sound wave propagation (triangles) tests, as measured by the $L_2$ error norm. The filled symbols show the second order scheme (equation \ref{Eqn:DragImplicit}) and the hollow ones show the first order scheme (equation \ref{Eqn:ExactLin}). When the spatial dependence is included (sound wave test), the accuracy is slightly degraded, but it is clear that equation \ref{Eqn:DragImplicit} remains second order.}
\label{Fig:tStepConv}
\end{figure}

\boldtxt{To demonstrate the accuracy of the time-stepping schemes, we show the $L_2$ error norm for the velocity in Fig. \ref{Fig:tStepConv} for two tests in which we have prescribed a total gas acceleration $a \equiv \grad{P}/\rho + a_g$. In the first we used a 1D forced oscillation, $a = V \omega \cos(\omega t)$ and we integrated both the gas and dust using the leap-frog schemes. For the second test we used a travelling wave $a = V \omega \sin(\omega t - k x)$, but integrated only the dust using the leap-frog scheme\footnote{The motivation for this is just that we want to ensure we can evaluate $v_g$ at the location of the dust. We show the full problem using interpolation and SPH forces in section~\ref{Sec:DustyWave}.}. The two tests are chosen to highlight the effect of neglecting the relative advection term, which is identically zero in the first example. For the parameters we have used $V = 10^{-4}$, $k=w =1$, and a
stopping time of $t_s = 0.05$. For the $L_2$ error norm, we use
\begin{equation}
L_2 = \left[ \frac{\sum_i (v_i - v_0(x_i,t))^2}{\sum v_0(x_i,t)} \right]^{1/2},
\end{equation}
where, for the forced oscillation we used the analytical solution for $v_0$, and a high accuracy numerical solution for the travelling wave. For the range of number steps shown, both the regimes $\Delta t \ll t_s$ and $\Delta t \gg t_s$ are tested, which corresponds to $N \gg 20$ and $N\ll 20$ for these parameters. We see that in both cases equation \ref{Eqn:ExactLin} produces a first order scheme, while equation \ref{Eqn:DragImplicit} is a second order scheme. 

If an explicit update for the dust velocity had been used, then for $N \lesssim 50$ the integration would have lead to significant error. However, as the stopping time decreases the power of our scheme becomes apparent, since we find essentially the same error regardless of stopping time with our scheme, and an accuracy of 0.1 per cent can be achieved with 100 steps. Many more steps would be needed for tightly coupled particles ($t_s < 0.01$) and an explicit scheme. One of the main attractions of our scheme is the ability to accurately reproduce the forces in poorly resolved flows, i.e. those with $h > c_s t_s$.} 

For non-linear drag laws expressions equivalent to equations \ref{Eqn:ExactLin} and \ref{Eqn:DragImplicit} cannot be expressed in terms of elementary functions, except in the case $\vec{a}_d = \vec{a}_g - \nabla P / \rho_g = 0$ \boldtxt{(although see \citet{Laibe2014a} for the case of quadratic drag laws)}. In this case the only options are explicit integration or an operator split approach. In protoplanetary discs the transition to non-linear drag regimes occurs for particle sizes ${a \gtrsim 6.6\,(M/M_{\rm MMSN})^{-1} (R/\unit{AU})^{2.5} \unit{cm}}$, where $M/M_{\rm MMSN}$ is the ratio of the disc mass to the minimum mass of the Solar nebula \citep{Johansen2014}. Since these particles have ${\rm St } > 1$, the time-steps are limited by the orbital velocity and $\Delta t \ll t_s$, which means that explicit integration is suitable.

Using analytical solutions means that the time-steps are not limited by the stopping time, and we can use the standard limits from the gas time-step. We find that a Courant-like condition is sufficient to make sure that $\vec{v}_g$ does not change too much over a time-step. We use a signal velocity
\begin{equation}
v_{\rm sig} = \max(| \vec{v}_d - \vec{v}_g|, c_s), \label{Eqn:Courant}
\end{equation}
and set the time-step via $\delta t_C = \chi_1 h / v_{\rm sig}$, with $\chi_1 \sim 0.1$. For the sound-speed, $c_s$, we use the sound speed interpolated at the location of dust particles. Since the Courant condition is proportional to resolution, the dust velocity converges for $h \rightarrow 0$, even for $t_s \rightarrow 0$.  Additionally, we limit the time-step based on the gravitational acceleration, ${\delta t_a = \chi_2 \sqrt{ h / |\vec{a}|}}$. The limits have been sufficient in all tests presented, however it may be necessary to include an additional limiter in the presence of strong shocks \citep{Durier2012}.

\section{Numerical Validation}

In order to validate the scheme, we present several simple tests based upon the \textsc{dustybox}, \textsc{dustywave}, settling, and \textsc{dustyshock} tests described in \citet{Laibe2011} and \citep{Loren-Aguilar2014}, evaluated in the test particle limit. Using the test particle limit means it is possible to compare the simulations to semi-analytical solutions, of which few are available for general problems. Additionally, we consider multidimensional test problems, which are more stringent tests of the codes robustness and performance in problems that are closer to real physical scenarios.

\subsection{\textsc{dustybox} for constant drag coefficients}

\begin{figure}
\centering
\includegraphics[width=0.47\textwidth]{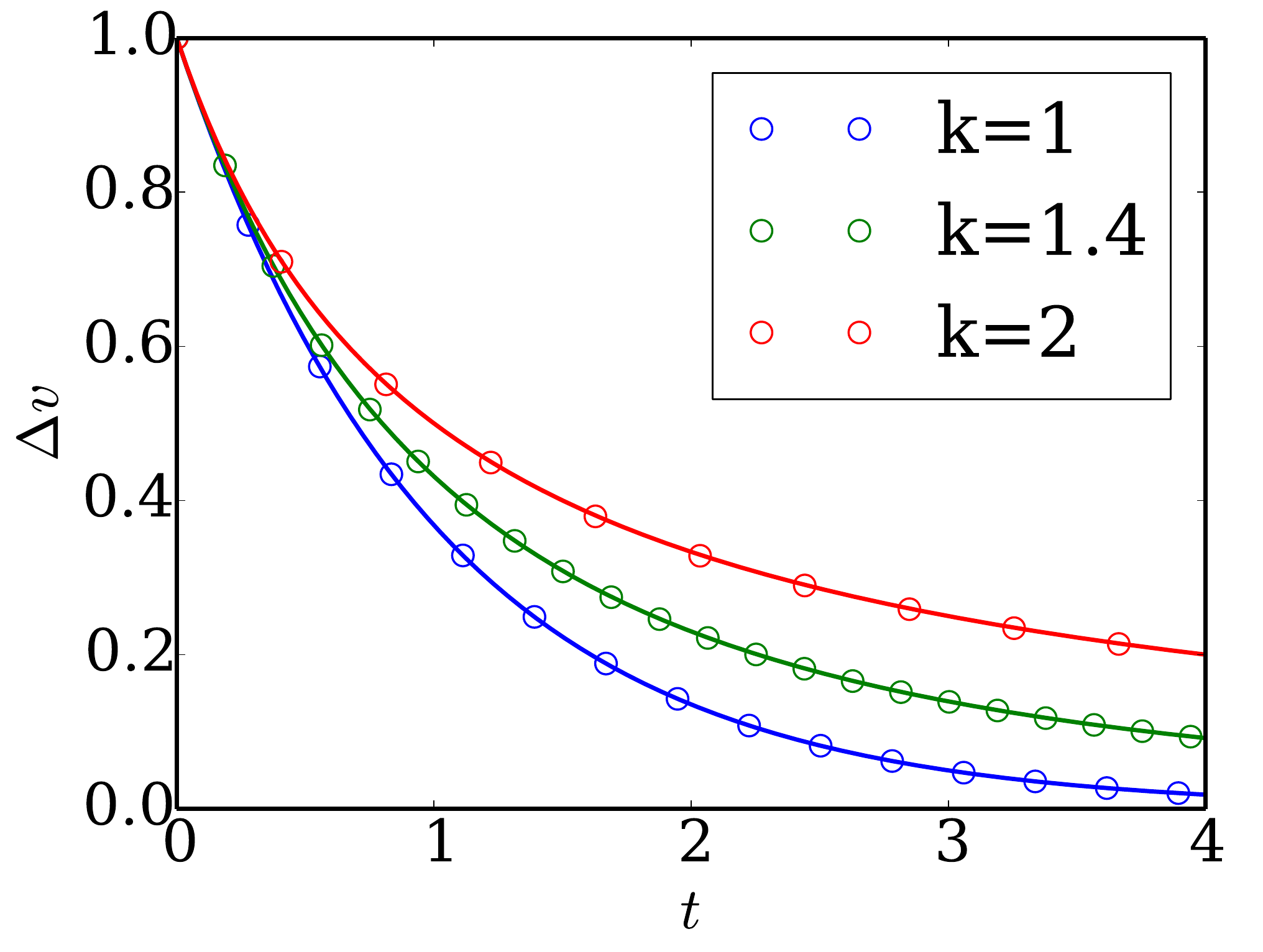}
\caption{Time evolution of the relative velocity between the gas and dust, $\Delta \vec{v} = \vec{v}_d - \vec{v}_g$, for power-law drag forces, $\vec{F}_d \propto - |\Delta \vec{v}|^k$, with varying index $k$. In each case the coefficients are set such that the stopping time $t_s = 1$ at $t = 0$. The SPH solution is given by the circles, while the analytical solution is given by the solid lines. The scheme is able to correctly reproduce the force, producing a peak velocity error of $3 \times 10^{-4}$.}
\label{Fig:StopT}
\end{figure}

The \textsc{dustybox} test is conducted in a 3D periodic box of uniform density at a resolution of $20^3$ gas particles placed on a Cartesian grid. We use $20^3$ dust particles, although since the motion of the dust particles are independent a single dust particle could be considered. The dust particles are placed on a grid offset from the gas. The initial dust velocity is taken to be $\vec{v}_d = 1$ in code units, and the gas velocity $\vec{v_g} = 0$.The particles are then evolved using the analytical solution time-steps for different drag regimes, which is possible for the non-linear drag laws since there are no external forces. In each case the drag coefficients are chosen such that $t_s = 1$ at $t = 0$. 

\boldtxt{This test should be trivially passed by our time-stepping scheme since the velocity update is exact in this case. Nevertheless, this verification is important since some early schemes struggle with even this simple test, although generally this can be solved with appropriate kernels \citep{Monaghan1995,Laibe2012a}.} As can be seen in Fig.~\ref{Fig:StopT}, the scheme successfully captures the drag forces in each regime, achieving a force error of less than 0.5 per cent, which verifies the accuracy of the interpolation scheme. The results are independent of the ratio of stopping time, $t_s$, to gas time-steps, $h / c_s$. Similar results are found using an explicit time-stepping scheme, albeit at significantly higher computational cost when $c_s t_s \ll h$.

\subsection{\textsc{dustywave}}
\label{Sec:DustyWave}

\begin{figure}
\centering
\includegraphics[width=\columnwidth]{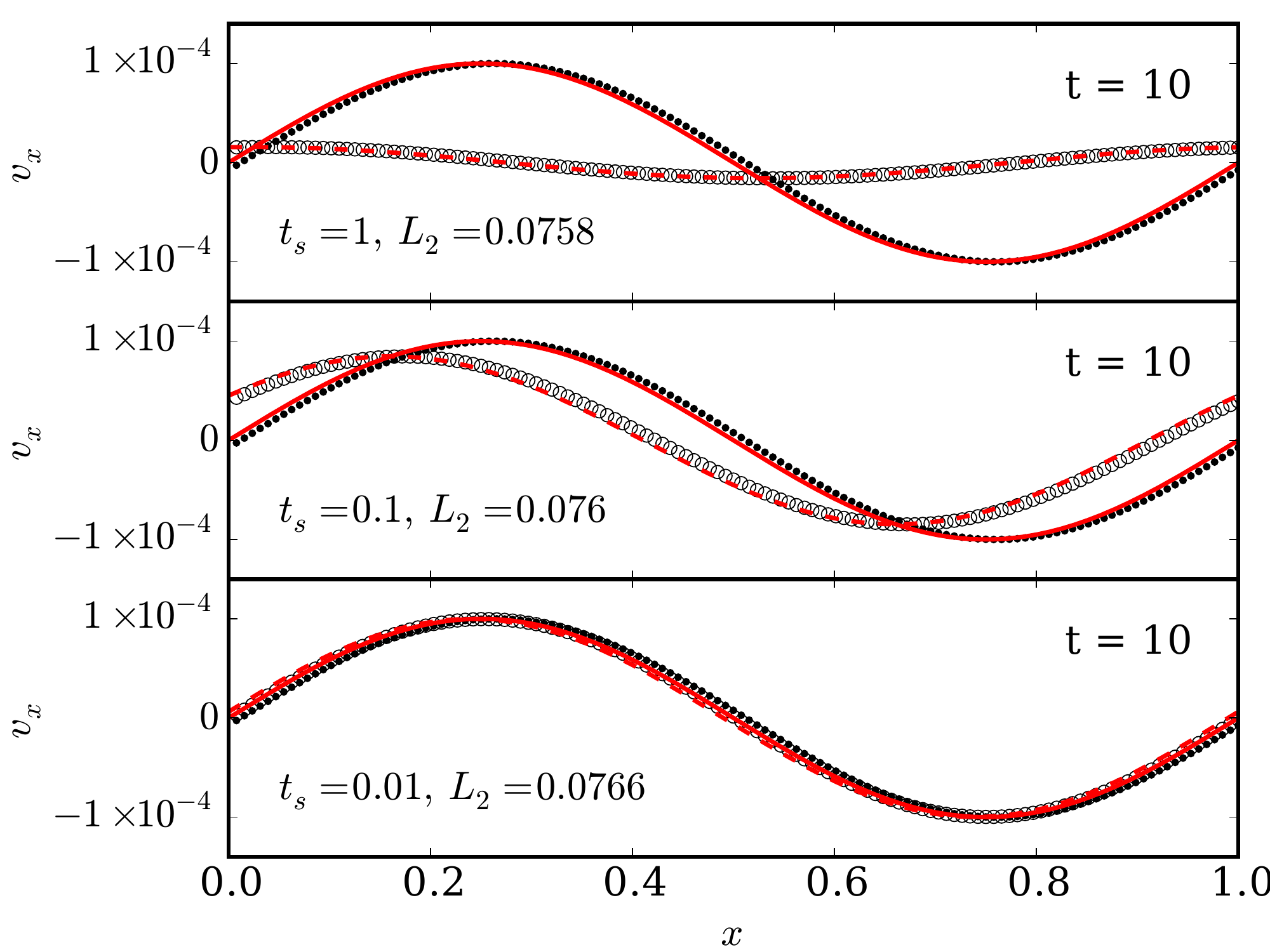}
\caption{\textsc{Dustywave} linear sound-wave test for various stopping times, $t_s$. The red lines show the exact solution for the gas (solid) and dust (dashed), while the black points and circles show the numerical solution for the gas and dust respectively, using 128 particles per phase. The results show good agreement between the numerical and exact solution. The $L_2$ error norms are the same for all stopping times, but larger than the idealized case using the exact gas velocity for the force. The explanation is that the SPH force evaluation is the dominant source of error.}
\label{Fig:DustyWave}
\end{figure}

\boldtxt{The \textsc{dustywave} test \citep{Laibe2011} is essentially the travelling wave test that we used to verify the order of the time-stepping scheme in section \ref{Sec:TimeInteg}, but using live SPH particles and interpolation to calculate the velocity at the location of the dust particles. We have computed the test in one dimension, using 128 particles per phase. We switched off the artificial viscosity since our aim is to study the motion of dust particles rather than viscous dissipation, which depends sensitively on the artificial viscosity implementation \citep[see, for example,][]{Cullen2010}. 

In addition to the velocity perturbation $\delta v /c_s = 10^{-4}$, we set up a density perturbation $\delta \rho / \rho = 10^{-4}$ by perturbing the initial particle positions according to the integrated density along the wave. We have used an adiabatic equation of state, with $\gamma = 5/3$, and set the entropic
function $A = P / \rho^\gamma$ to be a constant in order to produce $c_s = 1$ in the unperturbed flow. In each test we have set the stopping time $t_s$ to be a constant. 

The results of the live test are shown in Fig.~\ref{Fig:DustyWave}, which show good agreement between the analytical solution and the numerical results. Although not shown, we see similarly good agreement for arbitrarily small stopping times, which shows that our numerical scheme successfully captures the evolution of a time-dependent flow, even when $h < c_s t_s$ and $\Delta t > t_s$, which occurs for $t_s < 0.01$.  In these tests, we find that the $L_2$ error norm for the dust velocity is essentially equal to the error norm for the gas and is independent of stopping time. This is due to the dominant source of error coming from the SPH force evaluation giving rise to a slight bias in the gas velocity and effective sound speed. We note that the good agreement in the limit of small stopping times is aided by the fact that the \textsc{dustywave} test is much less stringent in the test particle limit, because unlike the full two-fluid case the sound speed is not modified by the presence of dust, and numerical dissipation does not occur, which can be an issue for $h < c_s t_s$ and large dust-to-gas ratios \citep{Laibe2014a,Loren-Aguilar2014}.
}

\subsection{Settling in the Epstein regime}
\label{Sec:Settling}
\begin{figure*}
\centering
\includegraphics[width=\textwidth]{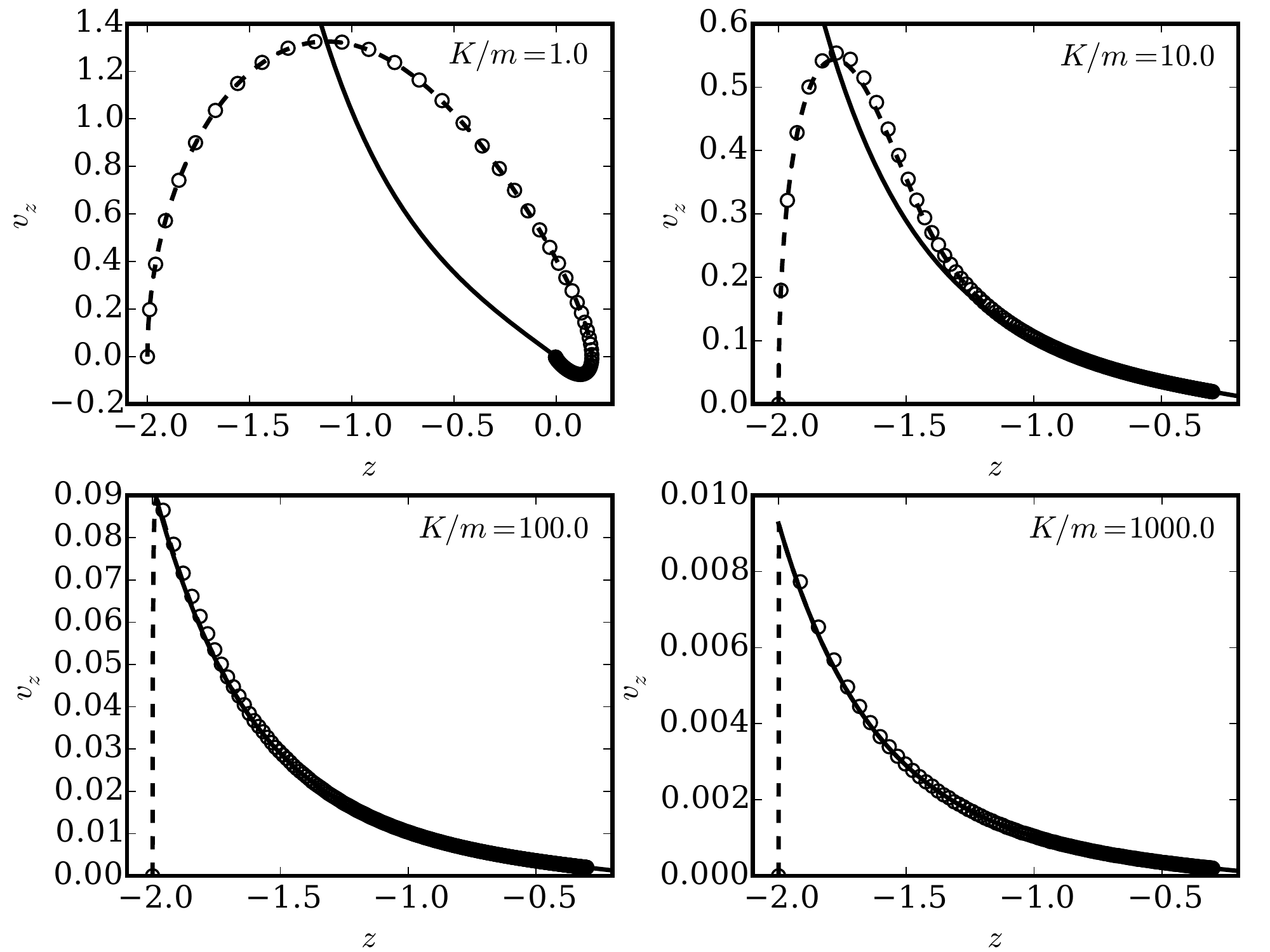}
\caption{Settling test for a range of drag coefficients, showing the evolution of the velocity of a particle starting from rest at $z=2$. The solid line shows the terminal velocity, while the dashed line show the exact solution for the particle velocity and the circles denote the numerical solution, which uses 100 particles per phase.  We see excellent agreement between the gas and dust and even for $K/m > 100$ we do not see signs of the spurious oscillations reported by \citet{Loren-Aguilar2014}.}
\label{Fig:Settling}
\end{figure*}

\boldtxt{This test involves the settling of particles under the action of gravity in a stratified density medium, and is useful for testing the scheme in the presence of a varying drag coefficient. For this test we mimic a protoplanetary disc in one dimension by applying an external force $a_g = a_d = - \Omega^2 z$. We fix the sound speed, $c_s$, and orbital frequency, $\Omega$, to 1. The gas is supported against gravity via pressure forces, which gives rise to a Gaussian density structure for the gas, 
\begin{equation}
\rho(z) = \rho_0 \exp\left( - \frac{z^2}{2 H^2} \right),
\end{equation}
where $H = c_s / \Omega = 1$ and $\rho_0 = \Sigma / \sqrt{2 \pi H}$. Following \citet{Loren-Aguilar2014}, we use 100 gas particles initially spread over the range $-2$ to $2$, with an initially constant density $\rho(t=0) = 1$, which implies $\Sigma = 4$, and relax the gas until it reaches equilibrium. Once the gas has reached equilibrium we introduced 100 dust particles, spread uniformly over the range $-2$ to $2$ as before. We mimic the Epstein regime by setting the stopping time $t_s = m / (K \rho)$, and start the dust particles from rest. 

In Fig.~\ref{Fig:Settling} we show the motion of a particle initially at $z=-2$ for a range of $K/m$. In all cases, the results show excellent agreement between the numerical solution and the exact solution, which for strong coupling tends quickly to the terminal velocity. We contrast our results with those reported by \citet{Loren-Aguilar2014}, who found spurious oscillations in the velocity for $K/m\ge 100$ at the same resolution. For $K/m = 100$, the stopping time $t_s \approx 0.05$ at $z = \pm 2$ and the smoothing length $h \approx 0.2$. For reasonable values of the Courant parameter, $\chi_1 = 0.1$, we find that the time-step, $\Delta t$, is of order the stopping time. We suggest the oscillations in their scheme may be due to the incorrect approach to the short-friction time limit for $\Delta t \gtrsim t_s$, rather than low number of particles. This is consistent with the disappearance of the oscillations at higher resolution, since the Courant condition enforces smaller time-steps and thus $\Delta t < t_s$. If our explanation is correct, then at every resolution there should be minimum stopping time below which their scheme produces oscillations. This test shows the benefit of using the full equation of motion for the velocity difference, rather than an operator splitting scheme.}

\subsection{\textsc{dustyshock}}

\begin{figure*}
\centering
\begin{tabular}{cc}
\includegraphics[width=0.47\textwidth]{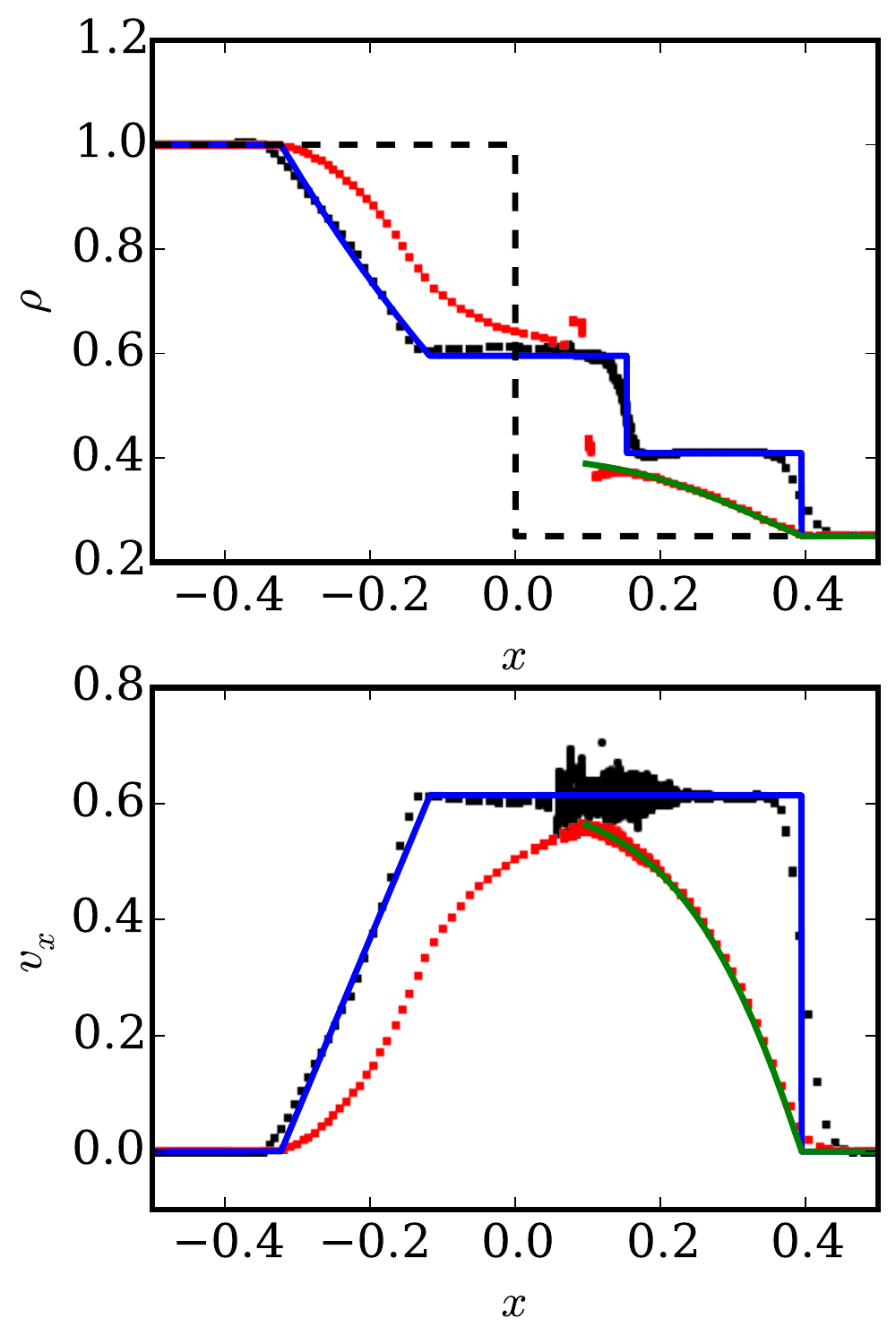} &
\includegraphics[width=0.47\textwidth]{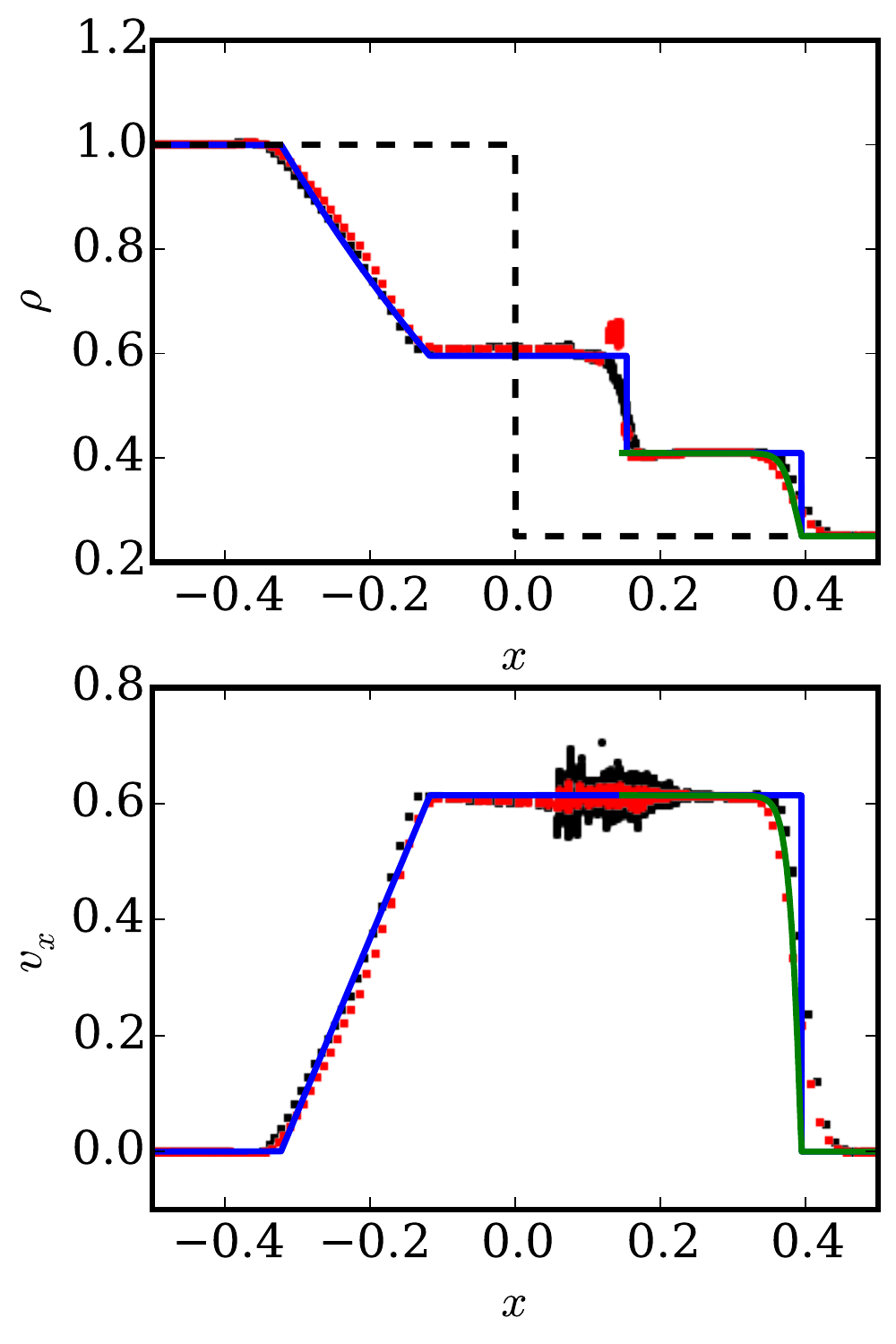}
\end{tabular}
\caption{Shock-tube tests for combined gas (black) and dust (red). The tests are calculated in 3D with stopping times $t_s = 0.1$ (left) and $t_s=10^{-2}$ (right). The analytical solution for the gas (blue) lines and  the dust initially to right of the contact discontinuity (green) are also shown. The choice of a cubic spline kernel and 64 neighbours gives rise to significant re-meshing noise in the gas velocity. However, the interpolated gas velocity at the location of the dust is considerably smoother, resulting in significantly less noise in the dust particles. Typically the velocity dispersion in the dust is less than 1 per cent of the sound speed, compared with 5 to 10 per cent for the gas.}
\label{Fig:ShockTube}
\end{figure*}

For the \textsc{dustyshock} test, we set up the initial conditions in 3D. The SPH particles are initially placed on a lattice with a spacing of 240 particles per unit length in the high-density region ($x < 0$), 152 particles per unit length in the low-density region ($x > 0$) and 64 neighbour particles. The masses and internal energies are set to give $\rho = 1$ and $P = 1$ for $x <0$ and $\rho =0.25$ and $P = 0.1795$ and the adiabatic index $\gamma = 5 / 3$. 

The dust particles are set up using the same density and number of particles as the gas. Even at arbitrarily low dust resolution the correct velocities are recovered, but an accurate density estimate requires a reasonable number of particles. An equal number of gas and dust particles has the advantage that the density in both phases experiences the same broadening, making a direct comparison simple.  Similarly to the \textsc{dustybox} test, the dust particles are placed on a lattice which is offset from the gas. The dust particles are then evolved using a linear drag law with a constant stopping time, $t_s$.

In the dust free case, the analytical solution for the gas is well known, which equally applies when dust is present in the test particle limit, \citep[see for example][]{Rasio1991}. In addition to the analytical solution to the gas, it is possible to compute the solution for dust via direct integration. For particles that initially have $x > 0$, the solution can be expressed parametrically in terms of the initial position of each dust particle, $x_0 = x(t = 0)$ and the time at which the shock reaches the particle $t_0 = x_0 / v_s$. For linear drag forces with constant coefficients this gives
\begin{equation}
v(x_0, t)  = \left\{ \begin{array}{l l}	
    0,	                                            & \quad x_0 \ge v_s t, \\
    v_m (1 - \exp [- (t - t_0)/ t_s]), & \quad x_0 < v_s t,
                    \end{array} \right .
\end{equation}
and 
\begin{equation}
x(x_0, t)  = x_0 + \left\{ \begin{array}{l l}	
	0,	 & \quad x_0 \ge v_s t, \\
	v_m	(t - t_0) - v(x_0, t) t_s & \quad x_0 < v_s t,
                    \end{array} \right .
\end{equation}
where $v_s$ is the shock velocity and $v_m$ is the velocity of the contact discontinuity. Since for a stationary shock $\grad\cdot (\rho \vec{v}) = 0$, the dust density can then be written in terms of the pre- and post-shock velocity, $\rho(x_0, t) = \rho_r v_s / (v_s - v(x_0, t))$, where $\rho_r$ is the initial dust density for $x > 0$.  For non-linear forces similar expressions can be derived straight-forwardly.

\begin{figure*}
\centering
\includegraphics[width=\textwidth]{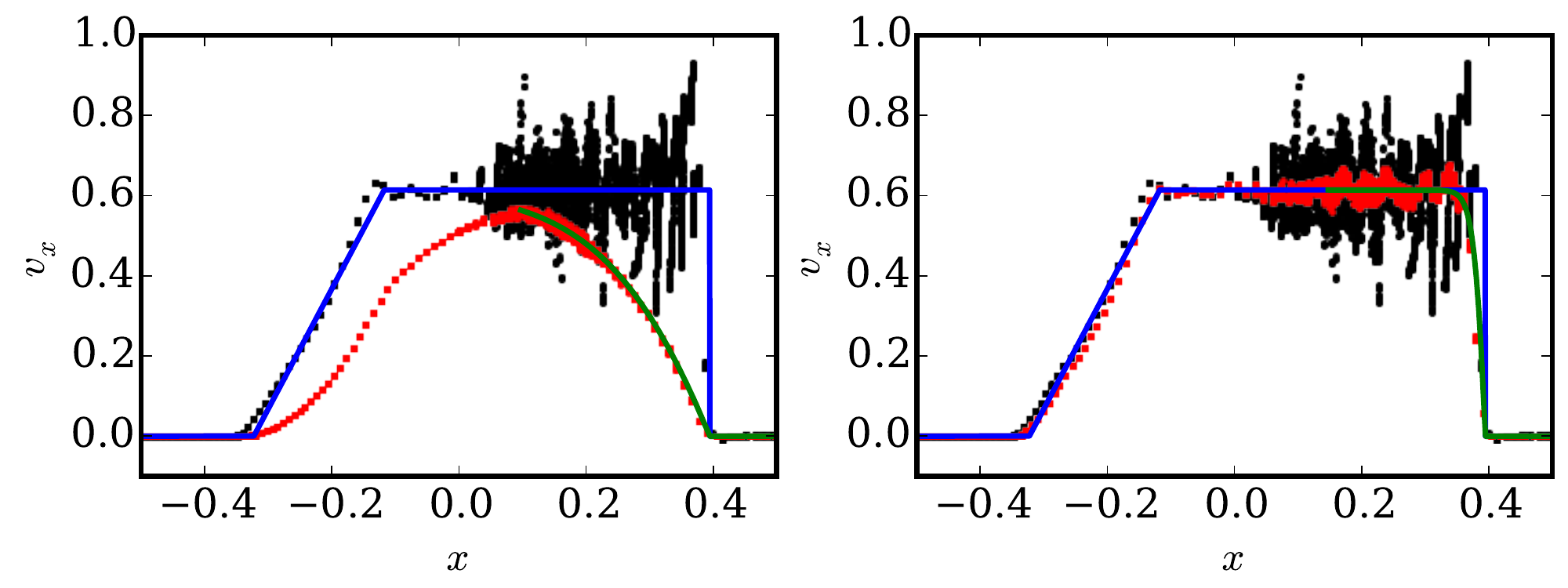}
\caption{Velocity in the shock-tube test with $t_s = 0.1$ (left) and $t_s =0.01$ (right) when a low viscosity parameter $\alpha = 0.1$ is used. The phenomenon of post-shock ringing in the gas is well known. While this translates into post-shock noise in the dust velocity, the amplitude is much smaller as the interpolation averages over the gas velocity.}
\label{Fig:ShockTubeLowVisc}
\end{figure*}

\begin{figure}
\centering
\includegraphics[width=0.47\textwidth]{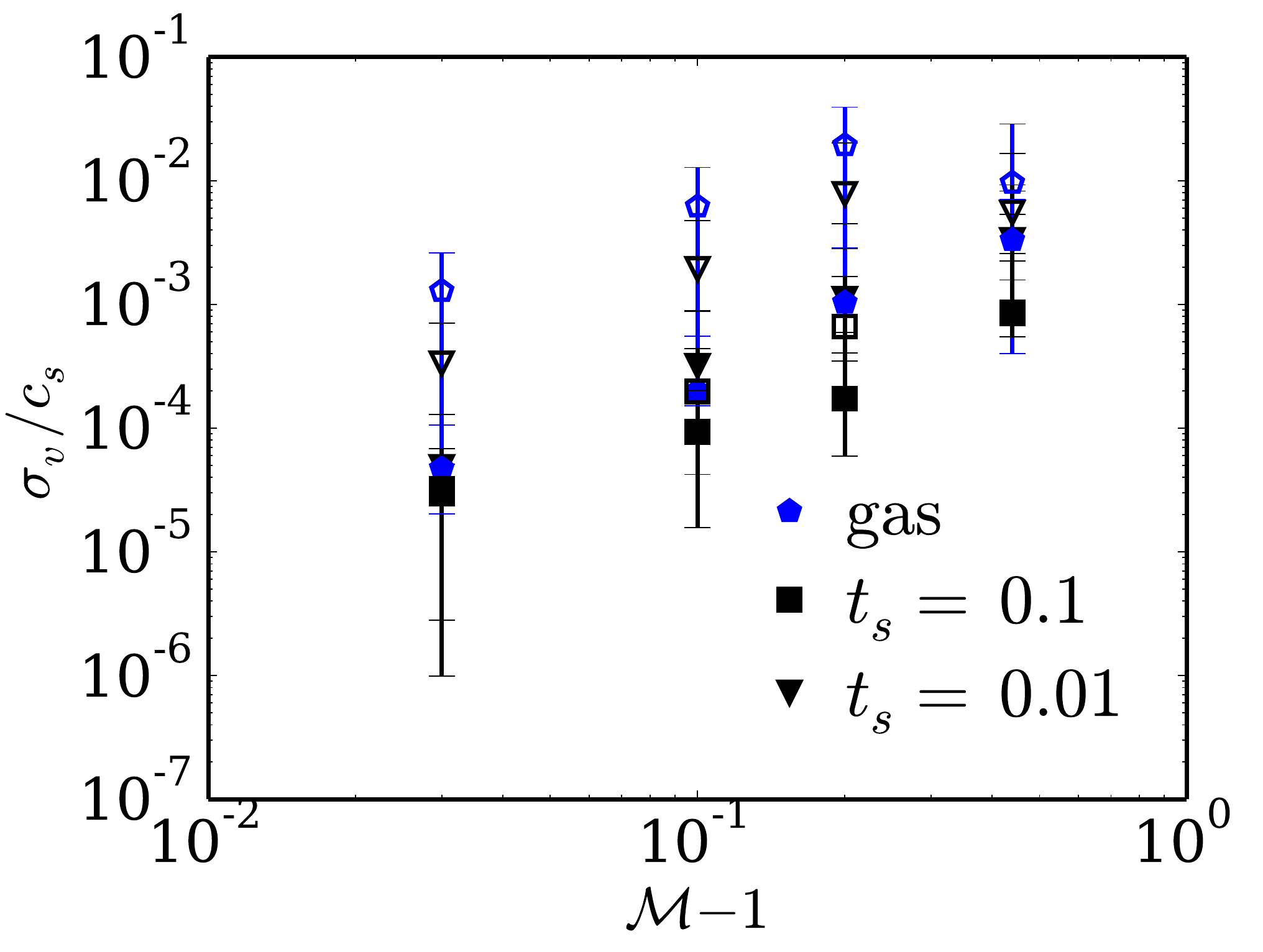}
\caption{Velocity dispersion in the post-shock gas in shock tube tests with varying Mach number, $\mathcal{M}$. The different symbols (triangles, squares, pentagons) refer to dust particles with differing stopping times, and the corresponding velocity dispersion in the gas. The filled symbols refer to standard simulations, whereas the hollow symbols show simulations where a low artificial viscosity, $\alpha = 0.12$, was used. The points show the median velocity dispersion measured and the error bars show the 25th and 75th percentiles. \boldtxt{We show $\mathcal{M} - 1$ on the x-axis to highlight the low velocity noise for weak shocks, as expected in protoplanetary discs.}}
\label{Fig:ShockTubeVelDisp}
\end{figure}

The results of the \textsc{dustyshock} test are shown in Fig.~\ref{Fig:ShockTube} for a linear drag law with stopping times, $t_s = 0.1$ (left) and 0.01 (right). There is good agreement between the analytical solution and the SPH solution. However, for moderate values of the stopping time, $t_s \approx 0.1$, it can be seen that the dust velocity is slightly larger in the simulations than the analytical solution. This arises due to the broadening of the shock and is most severe for the shortest stopping times. However, since the particles with the smallest stopping times reach the terminal velocity quickly the error in the position is bounded. For $h / t_s \lesssim |\Delta \vec{v}|$ we find the positional error is approximately $0.25h$, where $h$ is the gas smoothing length, but for longer stopping times the error is smaller. For the smooth post-shock flow in the \textsc{dustybox} problem the post-shock density is recovered. However, for more complex flows the positional error means that SPH simulations of gas and dust are limited to first order accuracy.

\subsubsection{Velocity noise and the consequences of too little artificial viscosity}

In SPH,  shocks compress the particle distribution anisotropically, which leads to re-meshing noise as particles re-arrange themselves to maintain a uniform distribution \citep{Price2012}. Using smoother kernels, such as the Wendland or quintic spline, can reduce this noise \citep{Price2012,Dehnen2012}. Since the re-meshing noise is driven by pressure forces,  the dust particles do not undergo re-meshing, although the drag forces introduce a velocity dispersion into the dust. To measure the velocity dispersion in the shock tube tests it is necessary to subtract off the gradient of velocity since the broadening of the shock introduces a slight bias into the velocity of the dust particles. To this end we use the linear-exact SPH gradient estimator \citep{Price2012}, which results in an estimate of $\sigma_v = 3 \times 10^{-3} c_s$ in the post-shock gas. For the dust particles with $t_s = 0.01$ the velocity dispersion is the same as the gas. For more weakly coupled dust the velocity dispersion is lower, with $\sigma_v < 10^{-3} c_s$ for $t_s = 0.1$.

While these noise levels are considerably lower than the velocity dispersion typically measured in simulations of self-gravitating protoplanetary discs, 
we do find that the noise is sensitive to the value of the artificial
viscosity parameter, $\alpha$. As described above, early simulations of self-gravitating discs, including the first with dust, employed $\alpha = 0.1$ instead
of the usually employed $\alpha = 1$. As a result, 
the simulations fail to generate sufficient entropy in shocks which gives
rise to oscillations in the velocity and density behind the shock. We illustrate this in Fig. \ref{Fig:ShockTubeLowVisc}, which also shows the considerable noise in the dust velocity driven by these oscillations. We note that in this case, subtracting off the gradient produces a considerably lower estimate for the velocity dispersion in the gas ($\sigma_v \sim 10^{-2} c_s$)  than if
we use the difference between the simulation and analytical solution for the estimation ($\sigma_v \sim 0.1 c_s$). For the dust the difference between the estimators is smaller, since the oscillatory structure is less well defined in the dust.

Since the shocks in protoplanetary discs are weak, we have measured the velocity dispersion generated by the re-meshing noise for Mach numbers between 1.03 and 1.44, which are shown in  Fig. \ref{Fig:ShockTubeVelDisp}. In this range, we find $\sigma_v \propto \mathcal{M} - 1$, reflecting the relationship between the compression in the post-shock gas and the velocity noise in the gas. The low velocity dispersion introduced in the dust particles by weak shocks $\sigma_v < 10^{-2}c_s$ shows that even though $\alpha = 0.1$ under-produces entropy in shocks, planar shocks are unlikely to be the dominant factor determining the velocity dispersion of dust in simulations of protoplanetary discs. In the following sections, we consider multidimensional tests to explore whether the level of noise produced in the shock tube tests is typical of more general problems.

\boldtxt{We note that the level of velocity noise reported is implementation dependent and reflects the levels of noise expected in a commonly used SPH implementation. In particular, the results will be sensitive to the choice of kernel, since smoother kernels that use a larger number of neighbours, such as the Quintic Spline or Wendland $C^6$ kernel, produce lower levels of noise in the gas velocity \citep{Price2012,Dehnen2012}. In cases where numerical noise from shocks needs to be minimised, these kernels are highly recommended since the dust particles lack the ability to regularize their particle distribution (see also section~\ref{Sec:Shock2D}).}

\subsection{Shocks in 2D}
\label{Sec:Shock2D}

\begin{figure*}
\centering
\includegraphics[width=\textwidth]{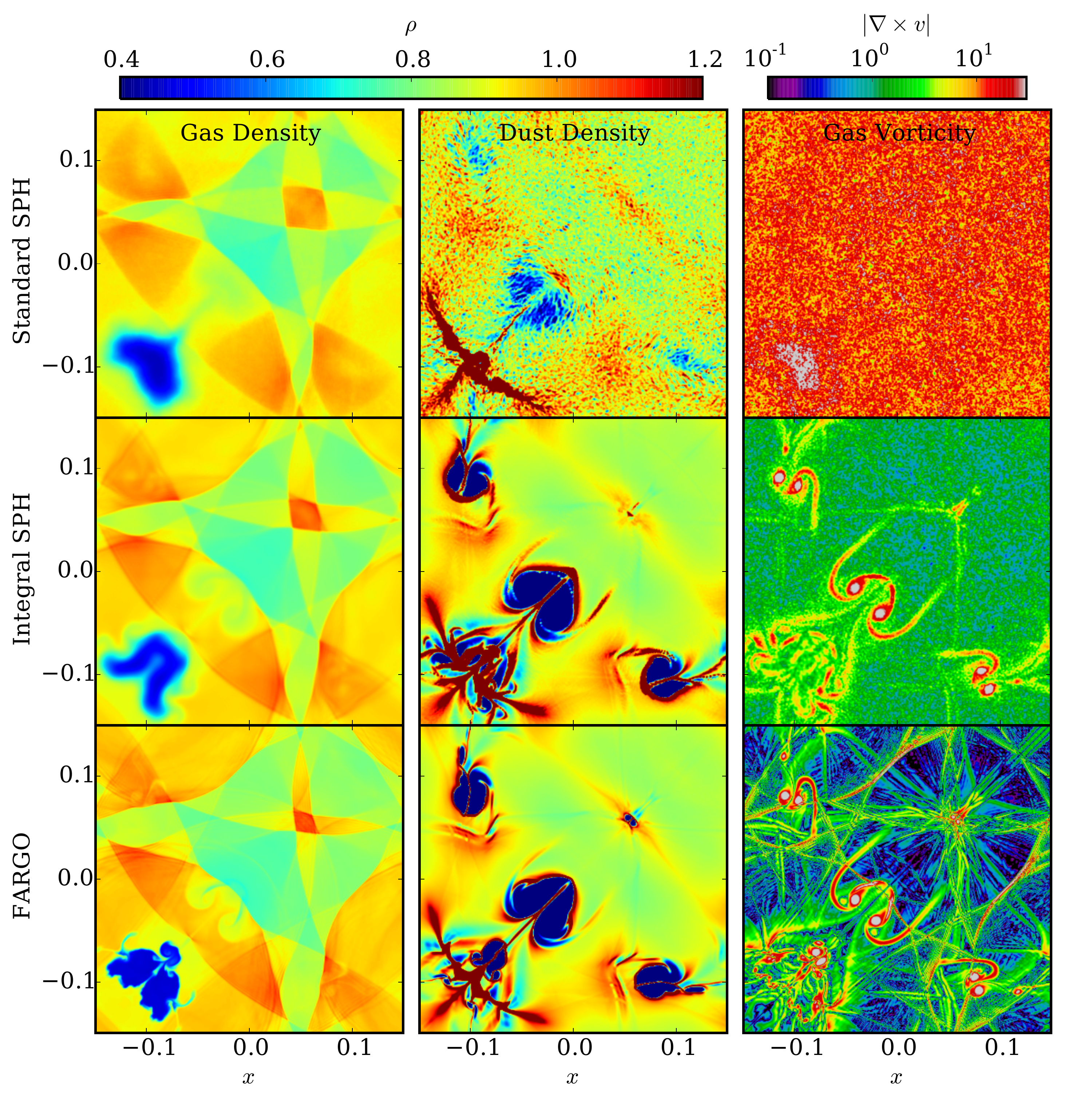} 
\caption{Gas and dust density (left \& middle) and gas vorticity (right) in the 2D shock problem at $t=0.7$. The top row shows the results for SPH using artificial conductivity and the \citet{Cullen2010} viscosity switch along with the cubic spline kernel. In the middle row a Wendland kernel and 50 neighbours were used, along with integral-based derivatives. In the bottom row, for comparison we show the results of the test problem as calculated by the grid-code FARGO, with the dust treated using a finite-volume approach. The noise in the gas density and velocity in the top panels has prevented the growth of Richtmyer-Meshkov instabilities, and is responsible for a large amount of noise in the dust density.}
\label{Fig:Shock2D}
\end{figure*}

The high symmetry of the Sod shock tube problem means that the uniformity of the initial conditions is preserved. However, in real applications the high symmetry of the initial conditions is broken as the system evolves. In order to show the effect that this has on the dust distribution we have conducted the implosion test \citep{Hui1999}, which we have modified to use periodic boundary conditions \citep{Sijacki2012}. For the initial conditions we consider a two-dimensional box with $-0.15 < x ,y< 0.15$.  The gas is initially at rest with density and pressure are $\rho =1$ and $P=1$, except for  $x + y < -0.15$, where $\rho = 0.125$ and $P = 0.14$, and the adiabatic index $\gamma = 1.4$. We have set up the test using equal mass particles and $Nx = Ny = 600$ particles in the high density region. The particles have been on a hexagonal lattice and we use 14 neighbour particles in 2D (equivalent to 39.4 neighbours in 3D). We add dust to this problem assuming a uniform initial density, using $Nx_g  = Ny_g = 600$ particles placed on a hexagonal lattice offset from the lattice of gas particles.  The stopping time is taken to be constant, $t_s = 0.1$.

The shocks launched by the initial discontinuities interact in the low density region and should drive a jet of gas along the symmetry axis $x = y$. The interaction of this jet with multiple shocks leads to Richtmyer-Meshkov instabilities. Vanilla SPH, as implemented in \textsc{gadget-2}, is known to struggle with such instabilities \citep{Agertz2007}. Although we are more interested in how the dust evolves in a problem without a high degree of symmetry, we have included an artificial conductivity as described by \citet{Price2012} and a viscosity limiter \citep{Cullen2010} to help reduce the damping of instabilities. 

We have also run the test using the grid-code FARGO \citep{Masset2000}, which is based upon the ZEUS algorithm \citep{Stone1992} to provide a reference solution for the dust. Dust has been included in FARGO using a finite volume approach in a similar way to \citet{Zhu2012}, using explicit time-stepping for the dust acceleration. Strictly a finite volume approach is not valid for this problem, since the dust velocity becomes multivalued in the region $x,y<0$. While this results in density artefacts that arise on the grid scale, this does not degrade the solution elsewhere, which can be confirmed by comparison with the SPH simulations. The FARGO simulations were run using $600 \times 600$ cells.

In Fig. \ref{Fig:Shock2D} we show the density at $t = 0.7$, after approximately a sound-crossing time. In the simulations in the top row, the large scale density structure is reasonably well reproduced, and the combination of a viscosity limiter with artificial conduction allows the growth of instabilities. However, the instabilities remain weak due to large amounts of noise in the density and velocity on scales of a smoothing length. The noise is particularly evident in the vorticity, which is largely featureless. The failing of standard SPH to reproduce the vorticity was noted by \citet{Sijacki2012}, and we find that while using a viscosity limiter and artificial conductivity helps the instabilities grow, it also results in a more noisy solution. The larger noise in the vorticity is associated with the lower viscosity - using a fixed $\alpha = 1$ not only damps the instabilities but also damps the velocity noise and helps maintain a regular particle distribution. The noisy gas velocity translates into considerable noise in the dust density on scales close to a smoothing length, even in regions that are not affected by the instabilities. The difference in the noise in the dust and gas densities provides a clear illustration of the importance of the self-regulation in SPH  in producing an accurate solution, since the dust has no mechanism by which it can reduce the noise in the density distribution once it has been introduced. 

Since the higher density noise in the dust is associated with the fact that the dust particle distribution is less regular, the noise could be decreased by increasing the number of neighbours. For a standard cubic spline kernel particle, the  pairing instability prevents the use of many more neighbours in the gas \citep{Price2012}. Since the drag forces have been included using averages of the gas velocity the dust particle should not be affected by the pairing instability. In this case it makes more sense to spend additional computational time to ensure the gas dynamics is computed accurately. To demonstrate this, we recompute the test using a modern SPH implementation that uses a Wendland kernel, 50 neighbours, and integral based derivatives. The implementation is essentially identical to the $\mathcal{F}_3$ formulation of \citet{Rosswog2014}. Increasing the number of neighbours results in an additional computational cost of a factor of three. Since the integral based derivatives are required for the \citet{Cullen2010} viscosity switch, they can also be used for the force calculation at negligible extra cost, although we note that this represents a 20 per cent increase over simulations in which the viscosity parameter is kept at a fixed value of $\alpha = 1$.

The results using this scheme are shown in the middle row of Fig \ref{Fig:Shock2D}. The effect of these modernizations on the gas density is relatively minor, resulting in sharper shocks and  stronger Richtmyer-Meshkov plumes. The sharper shocks are a result of  the reduced noise and give the appearance of higher resolution, although the same resolution has been used, albeit at considerable extra computational cost.  However, the improvement that this makes on the dust density is staggering. The density noise in smooth parts of the flow is significantly reduced and vortices in the gas appear that are devoid of dust. These results are in good agreement with the FARGO calculations, shown in the bottom panel of Fig. \ref{Fig:Shock2D}, which exhibit a higher effective resolution. The main difference arise in the region $x,y <0$, reflecting the both the higher resolution of the FARGO simulations in this region, along with the issues related to the dust velocity being multi-valued in this region. 

The vast improvement in the dust density distribution with the state of the art SPH calculation reflects the fact that reasonable agreement in the gas density can be achieved even when the gas velocity structure is dominated by noise. Since the gas velocity determines the dust dynamics, the dust density reflects the accuracy achieved in the velocity structure. The modern SPH implementation clearly does a much better job at reproducing the gas velocity, which also results in a much better reproduction of the instabilities. As this test shows, an accurate velocity structure is essential for reproducing the dust dynamics and convergence in the velocity structure can be considerably more difficult to attain.


\subsection{Static disc problem}

In addition to the simple verification tests above, we present a test for the code in a mode of operation that is close to the conditions present in protoplanetary discs. For this, we consider the motion of an isothermal gas in a fixed potential, consisting of an axisymmetric background potential with an imposed spiral perturbation. For a logarithmic potential the solution can be calculated along streamlines that orbit at a given average radius \citep{Roberts1969,Shu1973}. In particular, we follow the method of solution given in \citet{Gittins2004}. We use a logarithmic spiral perturbing potential, $V_s$, with a constant pitch angle, $i$, given by
\begin{equation}
V_s = A_0 R \exp(-\epsilon_s R) \cos( \chi),
\end{equation}
where 
\begin{equation}
\chi = - \frac{m}{\tan i}\ln(\epsilon_s R) - m(\theta - \Omega_p t).
\end{equation}
For the number of spiral arms, $m$, we use $m=2$, the pitch angle, $\sin i = 0.1$ and the pattern speed at which the spiral potential rotates, $\Omega_p = 0.522$.  The constants $A_0$ and $\epsilon_s$ set the strength of the potential and the length over which it decays, and $R$ is the cylindrical radius. The spiral perturbation strength can be set relative to the background potential, we take $\epsilon_s = 0.85$ and $F = 0.05$ at $R=1$, where 
\begin{equation}
F (R)= |\grad V_s| / |\grad V_0| = \frac{m A_0 R \exp(-\epsilon_s R)}{R^2 \Omega^2 \sin i}.
\end{equation}

For the background potential, a Keplerian velocity $\Omega(R) = \sqrt{G M / R^3}$ could be used. However, the streamline solutions cannot be found in regions close to corotation, where the background velocity is subsonic, ${R (\Omega - \Omega_p) \sin i \lesssim c_s}$. Since the spiral modes present in protoplanetary discs are close to corotation it is not possible to directly probe the regime present in protoplanetary discs using the streamline solutions \citep{Cossins2009}. Instead, we use a flat rotation curve in which the streamline solutions have been well studied in relation to spiral galaxies. Nevertheless, this allows us to test how the code handles shocks in a shearing environment. For the background potential we use the form and parameters from \citet{Gittins2004}, which we scale to dimensionless values. The velocity is given by
\begin{equation}
v(R) = v_{\rm max} \sqrt{F_b \epsilon_b R \exp(- \epsilon_b R) +
						 1 - \exp(-\epsilon_d R)},
\end{equation}
where we set the maximum rotation velocity $v_{\rm max}$ from $v(R) = 1$ at $R = 1$.The parameters $F_b$, and $\epsilon_b$ define the strength and size of the galactic bulge, which we neglect by taking $F_b = 0$. The scale of the disc is set via $\epsilon_d = 17/3$ and the sound speed of the gas is $c_s^{-1} = 27.5$.

The streamline solutions are calculated in coordinates that denote the position along, $\xi$, and between, $\eta$, the spiral arms. Choosing $\eta = \pi - \chi$ means that $\eta = 2n\pi$ for integer $n$ at the potential minima. Choosing $\xi$ to be perpendicular everywhere to $\eta$ gives
\begin{align}
\eta &= \frac{m}{\tan i} \ln(\epsilon_s R) + m ( \theta - \Omega_p t) + \pi \\
\xi  &= -m \ln(\epsilon_s R) + \frac{m}{\tan i} ( \theta - \Omega_p t).
\end{align}

The streamline solution can be calculated at a given radius under the approximation that the local angular velocity, epicyclic frequency, perturbation strength and sound speed are constant along the streamline. The streamlines are found by choosing a value $\eta_s$ for the sonic point and integrating forward and backwards along the streamline, whilst looking for a shock. Once a shock has been found $\eta_s$ is varied until a period solution is found. For more details on the exact method, see Appendix A of \citet{Gittins2004}. In addition to the equations for the velocity along and perpendicular to the streamlines, $v_\eta$ and $v_\xi$, we solve for the position about the disc by integrating
\begin{equation}
\deriv{\xi}{\eta} = \frac{v_\xi}{v_\eta}
\end{equation}
along a streamline \citep{Roberts1969}.

\begin{figure*}
\centering
\includegraphics[width=\textwidth]{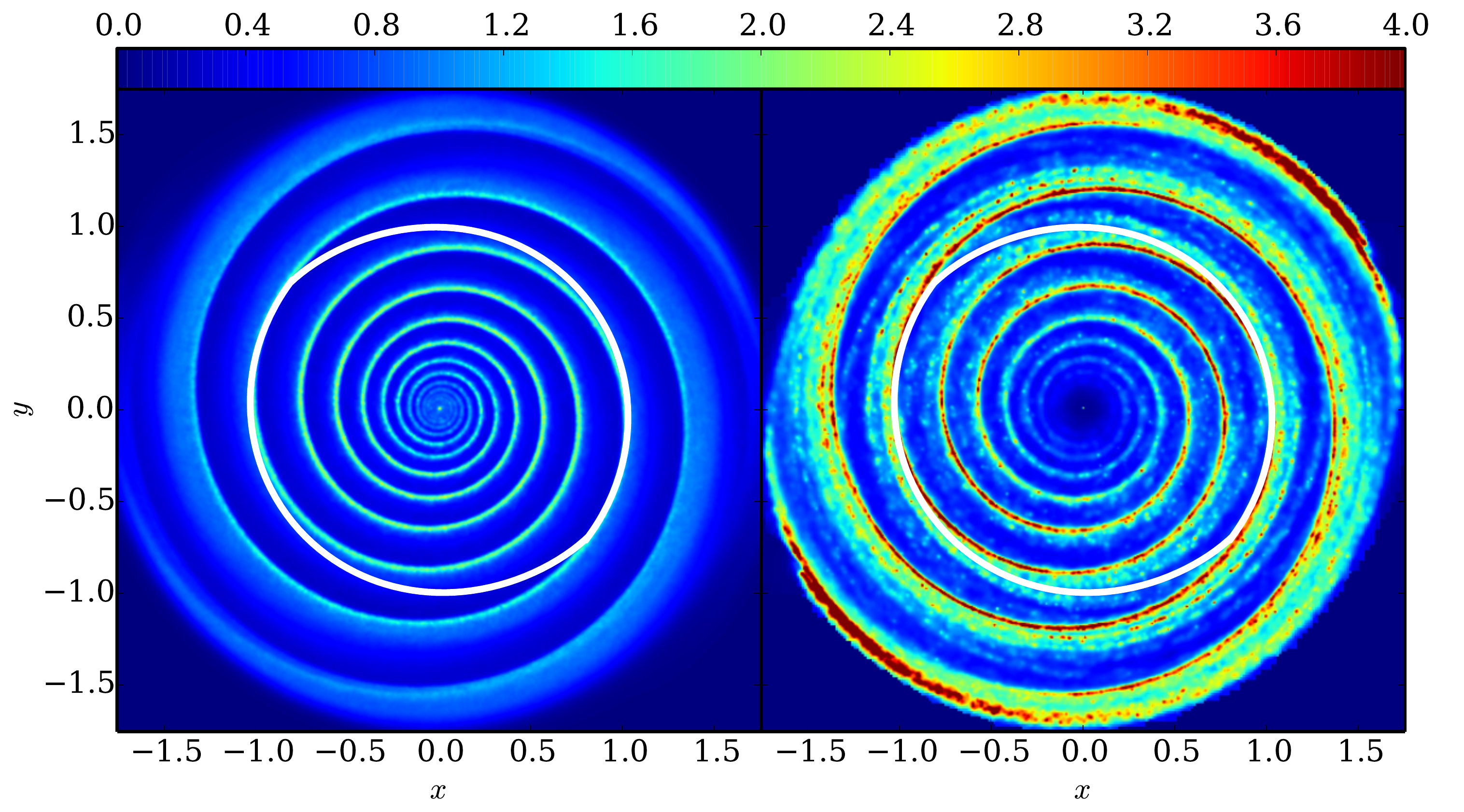}
\caption{Gas (left) and dust (right) density in the isothermal disc for the 3D simulation with $h/H = 0.16$ at $R=1$ and Stokes number, ${\rm St} = 1$. The dust density has been computed using twice the smoothing length of gas, to reduce the noise in the density estimate. The white line shows a gas streamline for comparison. The flow is anticlockwise and  the density maxima in the dust appears behind the density maxima in the gas.}
\label{Fig:DiscDensity}
\end{figure*}

\begin{figure*}
\centering
\includegraphics[width=\textwidth]{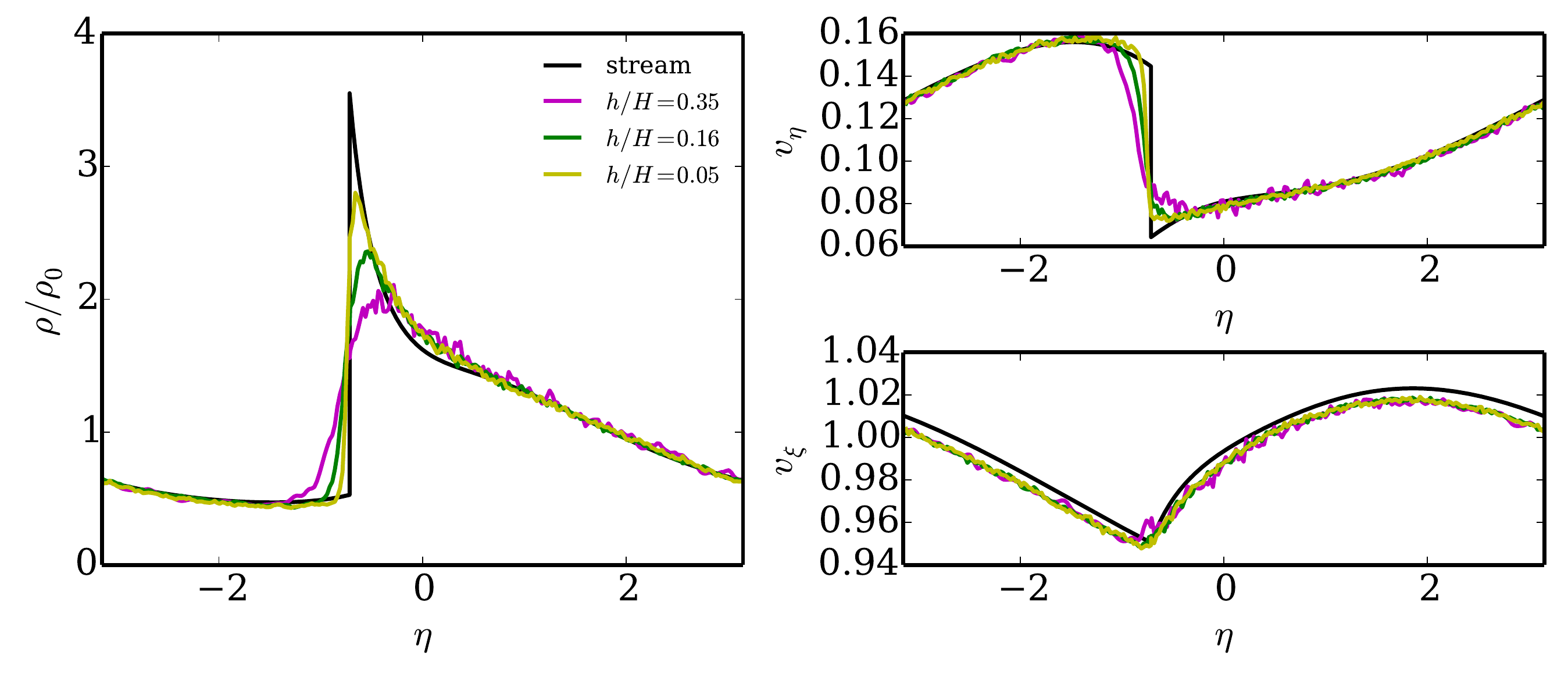}
\caption{Convergence of gas density and velocity perpendicular ($v_\eta$) and parallel ($v_\xi$) to the spiral equipotential lines at $R = 1$. The potential minima and maxima are denoted by  $\eta = 0$ and $\eta = \pm \pi$. The SPH simulations include two 3D simulations with resolution $h/H = 0.35$, and $0.16$ (magenta and green lines) and a 2D simulation with $h/H = 0.05$ (yellow). The black line gives the streamline solution. The SPH simulations converge well to the expected density and perpendicular velocity. However, the parallel velocity converges to a velocity that is 1 per cent lower than streamline solution.}
\label{Fig:GasStream}
\end{figure*}

\begin{figure*}
\centering
\includegraphics[width=\textwidth]{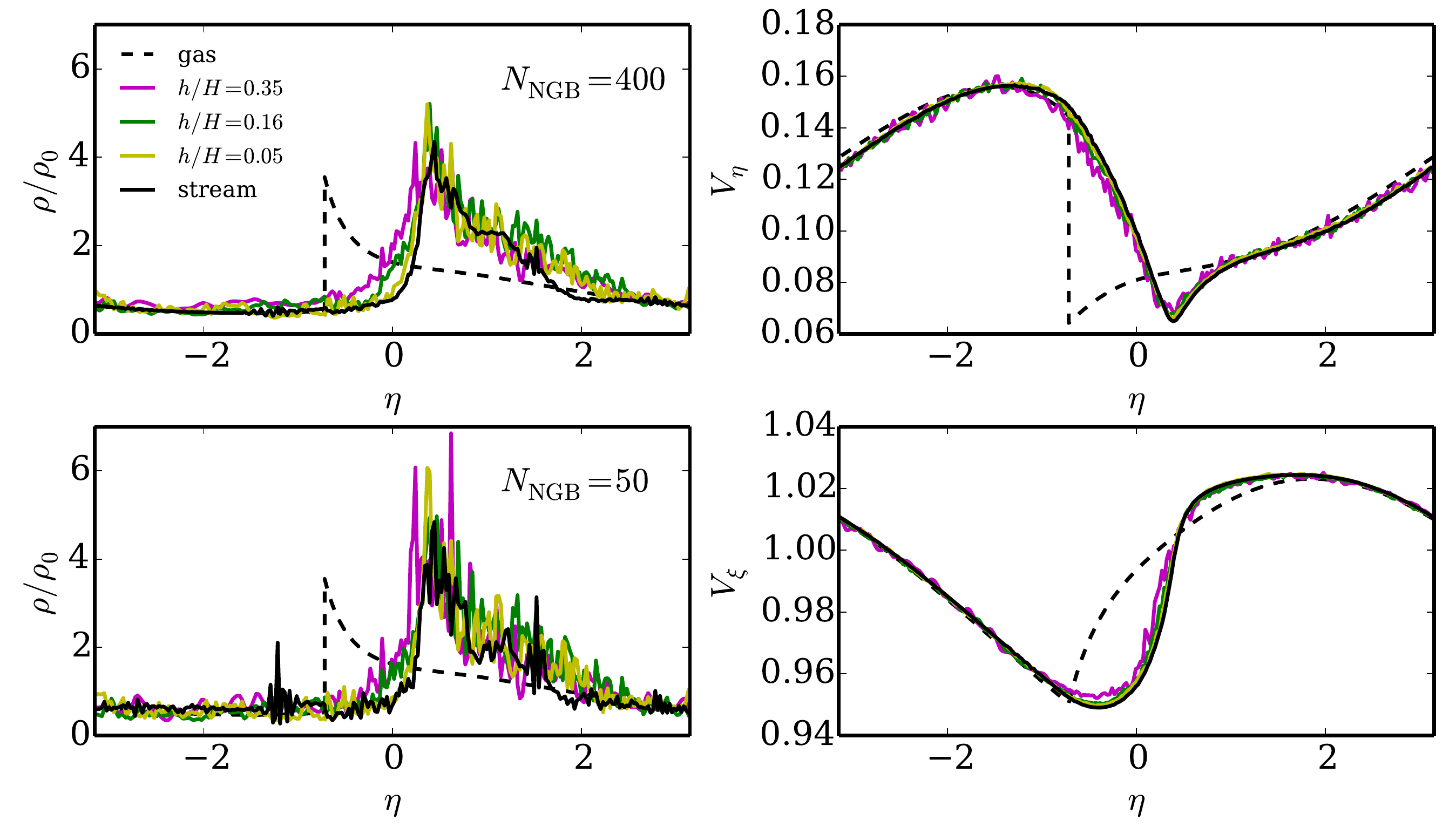}
\caption{Dust properties along a streamline for stokes number ${\rm St} = 1$. The black lines denote the gas and dust properties calculated from the streamline solution and the results from the SPH simulations are given by the coloured lines. The dust density has been calculated using the equivalent of 50 (bottom left) and 400 neighbours (top left) in 3D.}

\label{Fig:BoulderStream}
\end{figure*}

Once the streamline solutions have been calculated, the full structure of the gas can be built by interpolating between them.
However, the periodicity of the streamlines is only approximate, with the initial and final radii differing by roughly 1 per cent after the gas has travelled through an angle of $2 \pi / m$. For this reason instead of computing streamlines that give velocities that  are periodic over $\Delta \eta = 2\pi$, we look for solutions that are periodic over $\Delta \theta = 2 \pi / m$. This is motivated by the need for streamlines that do not overlap themselves in the interpolation process.

We now simulate the evolution of dust particles that evolve under the action of the spiral gravitational potential as well as drag forces exerted by gas whose density and kinematics is given by the analytic gas streamline solutions. Henceforth the dust properties derived from these simulations are described as `dust streamline solutions' and these are to be compared with the properties of the dust population obtained when the dust is instead introduced into a `live' SPH calculation
and the dust-gas drag computed as described above. For the `dust streamline solutions', up to $10^7$ particles are placed on a grid with an initial velocity taken to be the unperturbed background velocity. The particles are then evolved using a fourth order Runge-Kutta-Fehlberg integrator. Once all the dust positions are known, the dust density in any snapshot is then calculated using an SPH kernel sum.

The SPH simulations are conducted both in 2D and 3D. In the 3D simulations, periodic boundary conditions are used in the $z$-direction to allow direct comparison between the 2D and 3D results. This avoids the additional complications associated with stratified discs and the settling of dust into the mid-plane which arises within them. While the difference between velocity distributions in 2D and 3D is an interesting issue in its own right, it complicates the comparison to the analytical solution.

The SPH particles are set up using a uniform density glass, and the gas is evolved under the static potential until they reach a steady-state,  roughly 35 orbits at $R=1$. Once the gas has reached a steady-state the dust particles are introduced and the system is evolved for a further 35 orbits. Tests conducted in which the perturbing potential was slowly applied to the gas show no significant differences to tests in which the potential was switched on immediately. In all results presented the latter approach was taken. The uniform density means the inner regions of the disc are less resolved than the outer regions; therefore, it is instructive to quote the resolution in terms of $h/H$, where $h$ is the smoothing length and $H = c_s / \Omega$ is the pressure scale-height. For comparison at $R=1$, a $h/H$ of 0.35 and 0.17 translates into 2.5 million particles  with a box-height of $Z = 3$ and 6 million particles with a box-height of $0.75$. In 2D $h/H = 0.05$ can be achieved with 4 million particles. The number of particles used here is larger than the $\sim 10^6$ particles typically used in protoplanetary disc simulations \citep[c.f.][]{Rice2004,Cossins2009,Meru2012}, which reflects the fact the model disc is thinner $H / R \sim 10^{-2}$ than protoplanetary discs ($H/R \sim 0.1$), resulting in the need for considerably more particles to resolve the pressure scale height.

The density structure in the disc is shown in Fig. \ref{Fig:DiscDensity}, for the 3D simulation with $h / H = 0.16$ and ${\rm St} =1 $, upon which a typical streamline is marked for comparison. Each streamline crosses both of the spiral arms once. In calculating the dust density 400 neighbours have been used. The large number of neighbours needed in the dust density calculation reflects the fact that no self-regularizing forces apply to the dust. For a comparison of how noisy the dust density is with the normal $\sim 50 $ neighbours, see Fig. \ref{Fig:BoulderStream}. The flow is anti-clockwise and the dust density peaks significantly downstream of the shock, since the dust particles respond to the change in velocity over a stopping time. 

Fig.~\ref{Fig:GasStream} shows the gas density and velocity along a streamline at $R = 1$, showing that standard SPH is able to adequately reproduce the streamline solutions. While the SPH simulations accurately reproduce $v_\eta$, there is minor discrepancy in $v_\xi$ since the streamlines converge to a velocity that is systematically 1 per cent lower than the streamline solution. While pressure support can reduce the velocity below its Keplerian value, the disc is isothermal and initially has no density gradient. Once the spiral perturbation is applied a density gradient does arise, giving an order unity change in the density over the size of the disc. However, since the disc is cold and the ratio $H/R \sim 10^{-2}$, pressure support should only contribute at the $10^{-4}$ level.

 In some cases, artificial viscosity can be responsible for producing an incorrect disc structure in SPH simulations, via viscous angular momentum transport. For example, it has been shown that when a constant $\alpha = 1$ is used in the differentially rotating Gresho-Chan vortex problem that simulations do not converge to the correct inviscid solution, but appear to converge to a different solution \citep{Springel2010}. However, it is possible to show that artificial viscosity is not responsible for the velocity difference observed in our simulations. Firstly, in our test problem the spiral perturbing potential is responsible for driving the density gradient. Furthermore, the tests with low viscosity also show the velocity difference, as do tests in which a viscosity switch similar to \citet{Cullen2010} was used. Since the streamline solutions are only approximate, and the streamlines only close to within 1 per cent, the streamline solutions themselves may be the origin of the error. While the discrepancy is clearly greater than the noise in the simulations, the agreement with the streamlines is good enough that we can compare the dynamics of dust in the simulations to the streamline solutions.

In Fig.~\ref{Fig:BoulderStream}, the properties of dust for a linear drag law with ${\rm St} = 1$ are shown along the same streamline at $R=1$. The density has been calculated using both 50 and 400 neighbours, showing that a large number of neighbours is needed to achieve an accurate density estimate. Increasing the number of neighbours has a much more dramatic effect on the density estimate than increasing the resolution, which reflects that the local density estimate is subject to Monte-Carlo noise once the smooth density field present in the initial conditions has been sheared away. Conversely, the noise in the dust velocity is considerably lower, and can be clearly seen to be converging to the streamline solutions as the resolution increases.

\begin{figure}
\centering
\includegraphics[width=0.47\textwidth]{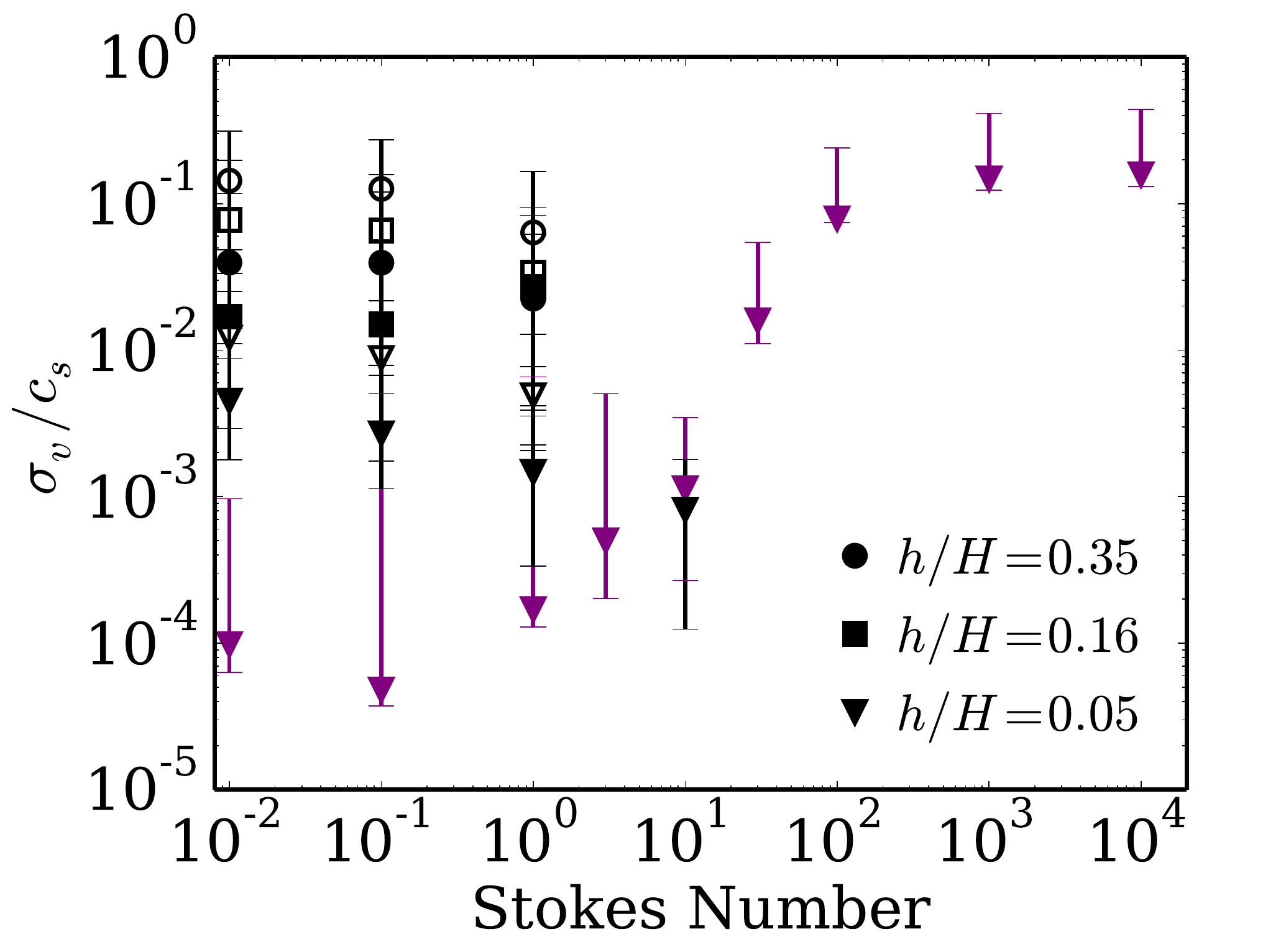}
\caption{Velocity dispersion as a function of Stokes Number for dust particles in SPH simulations (black) and streamline solution calculations (magenta). The different symbols mark the resolution of the simulations $h/H$. The sampling resolution of the streamline solution is close to the highest resolution SPH simulation. For the SPH simulations, filled and empty symbols mark simulations with standard viscosity and low viscosity respectively. The points mark the median velocity dispersion and the error bars denote the 25th and 75th percentile. The simulations at $h/H = 0.05$ were conducted in 2D, with both the higher resolution and lower number of degrees of freedom contributing to the lower noise.}
\label{Fig:BoulderVDisp}
\end{figure}

Since the velocity distribution of dust particles is fundamental to understanding the growth and evolution of dust grains in protoplanetary discs, it is important to check that the velocity distribution is not dominated by numerics. Since in a shearing disc there are variations in the velocity at the order of the sound speed, these need to be subtracted off in order to measure the velocity dispersion. To do this, we fit a third order polynomial to velocities at each point in space. We fit the velocity using a smoothing length that gives 400 neighbours to reduce the noise. We then subtract off the fitted velocity field and calculate the velocity dispersion from the residuals, again weighted by the kernel function. The method we use to fit the polynomial is identical to that described in \citet{Read2012}, except we fit a 3rd order polynomial. For more details, see Appendix C of \citet{Read2012}.

The velocity dispersion for the dust particles for a range of stopping times and resolutions is shown in Fig.~\ref{Fig:BoulderVDisp}. \boldtxt{The use of analytical solutions in the time-step means that we can investigate a range of Stokes numbers, ${\rm St} \lesssim 0.1$, that  would be inaccessible using explicit time integration.} Since the resolution varies as a function of radius, we calculate the velocity dispersion at points along a streamline at $R=1$, for both the simulations and the streamline solutions. The simulations find normalized velocity dispersions in the solid component of as little as $\sigma_v / c_s = 10^{-3}$ at the highest resolution and ${\rm St} = 1$, which is $3 \times 10^{-5}$ of the orbital speed. However, since this level of velocity is larger than that measured from the dust particles in the streamline solutions at the same resolution, it can still be considered numerical noise. For ${\rm St} = 10^{-2}$ even the velocity dispersion from the streamline solutions is dominated by numerical noise.

For the worst case scenario, the lowest resolution simulation with $\alpha = 0.12$, the median velocity dispersion in the gas particles reaches $0.1 c_s$. Typically the low viscosity simulations produce 5 times more velocity noise than those run with $\alpha = 1.2$. Similarly to the shock tube test, the velocity dispersion in the dust is typically a factor of 10 smaller than that in the gas. Comparison of the velocity dispersion measured from the simulations and streamline solutions near $h / H = 0.1$ shows that the velocity dispersion in the simulation is dominated by noise since the velocity dispersion of the streamline solutions is considerably lower. However, the velocity dispersion in this problem is orders of magnitude lower than that seen in simulations of the time-dependent structures in self-gravitating protostellar discs, which have $\sigma_v \sim c_s$ \citep{Rice2004}. Since the numerical noise is at the $10^{-2}c_s$ level in our $h/H = 0.16$ 3D calculation,  the velocity dispersion found in simulations of dust evolution is more likely to be associated with the physics of the problem, (i.e. the fluctuating spiral structure) than being a simple artefact of the numerical implementation. 

We have tested how using a modern SPH implementation affects the gas and dust dynamics in the spiral disc test. Although these methods help enormously with the two-dimensional shock tube test we found only a minor improvement when using integral-based gradients in the spiral disc problem. This can be understood since in the simple test cases the Wendland kernel and integral gradients produce a very regular particle distribution. However, in the static disc problem the shear prevents this from happening, which limits the benefits these improvements can provide. Additionally, we have run simulations using a \citet{Cullen2010} limiter with $\alpha_{\rm max} = 1.2$, and $\alpha_{\rm min} = 0$. With these parameters we found that $\sigma_v$ is 50 per cent higher than when using a fixed $\alpha = 1.2$, since viscosity acts to damp the noise in the particles. However, using a low fixed $\alpha = 0.1$ results in a $\sigma_v$ is several times higher than is produced when using the limiter. Clearly using a \citet{Cullen2010} limiter is preferred when in problems where it is necessary to reduce viscosity away from shocks.

\section{Discussion}

For a over a decade SPH has been relied on to provide insights into the dynamics of gas and dust in self-gravitating discs. Indeed to date, most of the global 3D simulations of self-gravitating discs and been performed with SPH (although \citet{Boley2010} investigated enrichment of solids in gas giants that are formed by gravitational instability using a 3D Eulerian code). It is therefore necessary to take a critical look at the fidelity of the method. While many code comparisons exist in the case of purely gaseous discs \citep{Meru2011,Meru2012,Paardekooper2011}, the dust implementation has been hitherto untested in this context \citep[although, for studies of dust in general see][]{Laibe2011, Laibe2012a, Ayliffe2012,Loren-Aguilar2014}.	

A key question in the case of self-gravitating discs concerns the velocity dispersion of solid material in the disc, since this controls the prospects for gravitational collapse in the dust phase as well as for particle growth or destruction. Previous simulations \citep{Rice2004,Rice2006,Gibbons2012} found a large velocity dispersion in the dust (of order a sound speed), which is unfavourable to either grain growth or gravitational collapse. It is therefore important to test whether this result is likely to be a numerical artefact.

In order to test the reliability of such codes, we have implemented dust particles into the SPH code \textsc{gadget-2}. We have used a two fluid approach and limited the study in the test particle limit where the feedback from the dust onto the gas can be neglected, allowing a direct comparison between simulations and analytical models. The code has been designed to accelerate the time-step using analytical solutions for the drag forces, in order to cope with particles that have Stokes number ${\rm St} \ll 1$. The chosen method has the advantage that it captures the full phase-space information of the dust, which is essential for evaluating the growth and survival of dust grains in protoplanetary discs. The code can handle the full range of Stokes numbers and produces the correct terminal velocity for $\Delta t \gg t_s$. We have focussed on extensively testing the code and its ability to tackle the questions in the dynamics of gas and dust in self-gravitating discs. \boldtxt{In many of the tests presented we have focussed on the dynamically interesting cases in which the stopping time is comparable, or slightly smaller than the dynamical time scale. The result is that our time-stepping scheme only really saves computational time for the shortest stopping times. However, demonstrating that the time-stepping scheme can handle both situations readily is important for realistic physical problems, such as protoplanetary discs in which both short and long stopping times are present.}

The code performs well in the standard \textsc{dustybox} and \textsc{dustyshock} test problems, in which the high symmetry means that the orderly distribution of the particles is maintained by the symmetry in the problems. \boldtxt{We also find good agreement in \textsc{dustywave} test, and have shown that our method is able to correctly reproduce the velocity of particles in a settling test, even at low resolutions and in strong drag regimes.} We also run the two dimensional shock-tube or implosion problem \citep{Hui1999}, which  involves interacting shock waves and is a much more severe test of a code's robustness. We have shown that correct modelling of the dust requires accurate modelling of the gas dynamics; Fig. \ref{Fig:Shock2D} contains a striking demonstration that a correct modelling of the gas density distribution does {\it not} guarantee that the gas dynamics (and the density distribution in the dust) is well represented since the dust is quite sensitive to noise in the gas velocity. In agreement with the results of \citet{Dehnen2012} in the case of the Gresho-Chan vortex test, we find that using a Wendland kernel (which  allows the number of neighbours to be increased beyond the traditional limit imposed by the pairing instability) makes a huge difference to the accuracy of the velocity field, as does using integral based derivatives. Since grid codes compare much more favourably in the two-dimensional shock tube problem, grid-based implementations of gas-dust mixtures will not show the same poor performance that standard SPH does in this problem. However, it would be interesting to see how one-fluid SPH codes perform in the presence of interacting shocks.

We also tested the code on a problem that includes spiral shocks in a shearing environment, which is applicable to shocks in self-gravitating discs, or structures that form from planet-disc interactions. We find that the code is able to reproduce accurately the motion of the dust particles, giving the expected solutions. However, in general this is less true for dust density. While the gas density in SPH calculations is accurate to the 1 per cent level for roughly 50 neighbours,  we have found that even with 400 neighbours 10 per cent accuracy in the dust density is not achieved. The difference arises because the SPH particle distribution is continually regularized by pressure forces, but the dust distribution is not. Once the disc has evolved for a sufficient length of time that the initial structure of the disc has been sheared away, the dust particles then randomly sample the density distribution. This means the density error, $\sigma_\rho \propto N_{\rm NGB}^{-1/2}$, for the dust particles while  $\sigma_\rho \propto N_{\rm NGB}^{-1}$ for the gas. Our results agree with the findings of \citet{Zhu2014}, who find that roughly 1000 neighbours are required to calculate the density with an accuracy of 10 per cent when particles are placed randomly in a volume. The reason this density noise is not seen in the shock tube problems is that the turbulence has not sufficiently deteriorated the initial particle distribution.

Since the dust density does not enter the equations of motion within the test particle limit, the error does not affect the ability of the code to calculate the motion of dust particles. The low velocity dispersion of the dust in the steady disc problem $\sigma_v \lesssim 10^{-2}c_s$ verifies this. The good performance of the code in the static disc test problem is promising, producing much lower velocity dispersion than seen hitherto in simulations of self-gravitating discs. However, there are some caveats. Firstly, using a low viscosity parameter, $\alpha = 0.1$, fails to generate enough thermal energy at shocks, which results in post-shock oscillations in the gas when an adiabatic equation of state is used. As the static disc test case is isothermal the energy generation requirements do not apply, which may help to reduce the noise generated. However, in all our test cases the amplitude of the velocity dispersion is smaller than the velocity jump across the shock $\sigma_v /c_s < \mathcal{M}$, setting a natural limit on the amount of noise introduced by shocks. It remains to be seen how the dusts dynamical state will be affected by these various code choices in the self-gravitating case and we leave this exploration (which will help to distinguish between physical and numerical effects) to a future paper.

Furthermore, when the dust is sufficiently dense to affect the dynamics of the gas, reducing the density error, which can be of order unity, will be important for producing reliable results. This is one area in which using a single-fluid approach would help, since the dust properties are evaluated at the location of the SPH particles and the $\sigma_{\rho_d} \propto N_{\rm NGB}^{-1/2}$ noise does not apply. However, in the presence of velocity noise the solution obtained by one-fluid approaches may also be deteriorated. Furthermore, in problems where the full phase space information is required, using a two-fluid approach with a large number of dust particle neighbours may be necessary to ensure an accurate force estimate. Indeed, similar results were reported by \citet{Laibe2012a}, who find significant noise in the dust density in a protoplanetary disc simulation, which led to artificial clumping. They found that using the gas smoothing length partially resolves the issue, since the gas has lower density and a larger smoothing length. We suspect that both error in the dust density and requirement that the gas is resolved on length scales present in the dust distribution are important for resolving the issue \citep{Ayliffe2012}.

Reducing the noise in the dust density distribution is even more important when dust self-gravity is important, since spurious over-dense clumps may collapse and feed back on the gas phase fragmentation. When self-gravity is important it is likely to be essential that a large softening length is used. Therefore, care must be taken in simulations of fragmentation in the dust layers of self-gravitating discs to avoid numerical artefacts. This is likely to apply to all particle based implementations of dust, independent of whether the gas is calculated using SPH or with a grid code.

\section{Conclusions}
We have implemented dust particles in the Smoothed Particle Hydrodynamics code \textsc{gadget-2}. Similarly to \citet{Loren-Aguilar2014} we avoid the need for small time-steps when the dust and gas are tightly coupled. However, unlike previous methods, our time stepping scheme produces the correct terminal velocity even when the time-step is much greater than the stopping time. We have applied a number of simple test problems that verify the code produces the expected results. The tests show the importance of producing an accurate velocity structure for capturing the dust dynamics, since a noisy velocity field introduces noise into the dust density that persists for the duration of the simulation.

We have used a rigidly rotating potential to test the dynamics of dust in spiral discs, and find that as long as the gas is well resolved numerical noise contributes a velocity dispersion $\sigma_v \sim 10^{-3}$ to $10^{-2} c_s$, below the levels found in simulations of {\it self-gravitating} protoplanetary discs. We find the dust density in disc problems is more difficult to estimate as shear destroys the high accuracy present in the initial density field. This means that the dust density is subject to Monte-Carlo noise and more than 400 neighbours are required for an accurate density estimate. While errors in the dust density do not affect the tests presented here, which are calculated in the test particle limit, when the dust density is high enough to be dynamically important, either through drag forces or self-gravity, care must be taken to ensure accurate results.

From these numerical experiments we have shown that for simulations in which the length scales in the gas are well resolved, $h/H \sim 0.1$, the dynamics of dust can be accurately modelled. These tests have been conducted in the test particle limit and in relation to SPH, but we expect them to hold more generally with the caveat that if the dust reaches high enough densities to affect the gas dynamics then a large number of neighbours is needed to ensure robust results. As long as care is taken to ensure the density noise is controlled, then simulations should be able to tackle some of the key questions related to the physics of dusty gasses.

\section*{Acknowledgements}

  We thank Giuseppe Lodato for useful discussions relating to this work, which has been supported by the DISCSIM project, grant agreement 341137 funded by the European Research Council under ERC-2013-ADG. \boldtxt{We would also like to thank the anonymous referees for many constructive points that helped to improve the paper.} 

\footnotesize{
  \bibliographystyle{mn2e_long}
  \bibliography{boulders}
}

\appendix
\section{Kernels \& Interpolation}
\label{Sec:AppendImpl}

\begin{figure}
\centering
\includegraphics[width=0.47\textwidth]{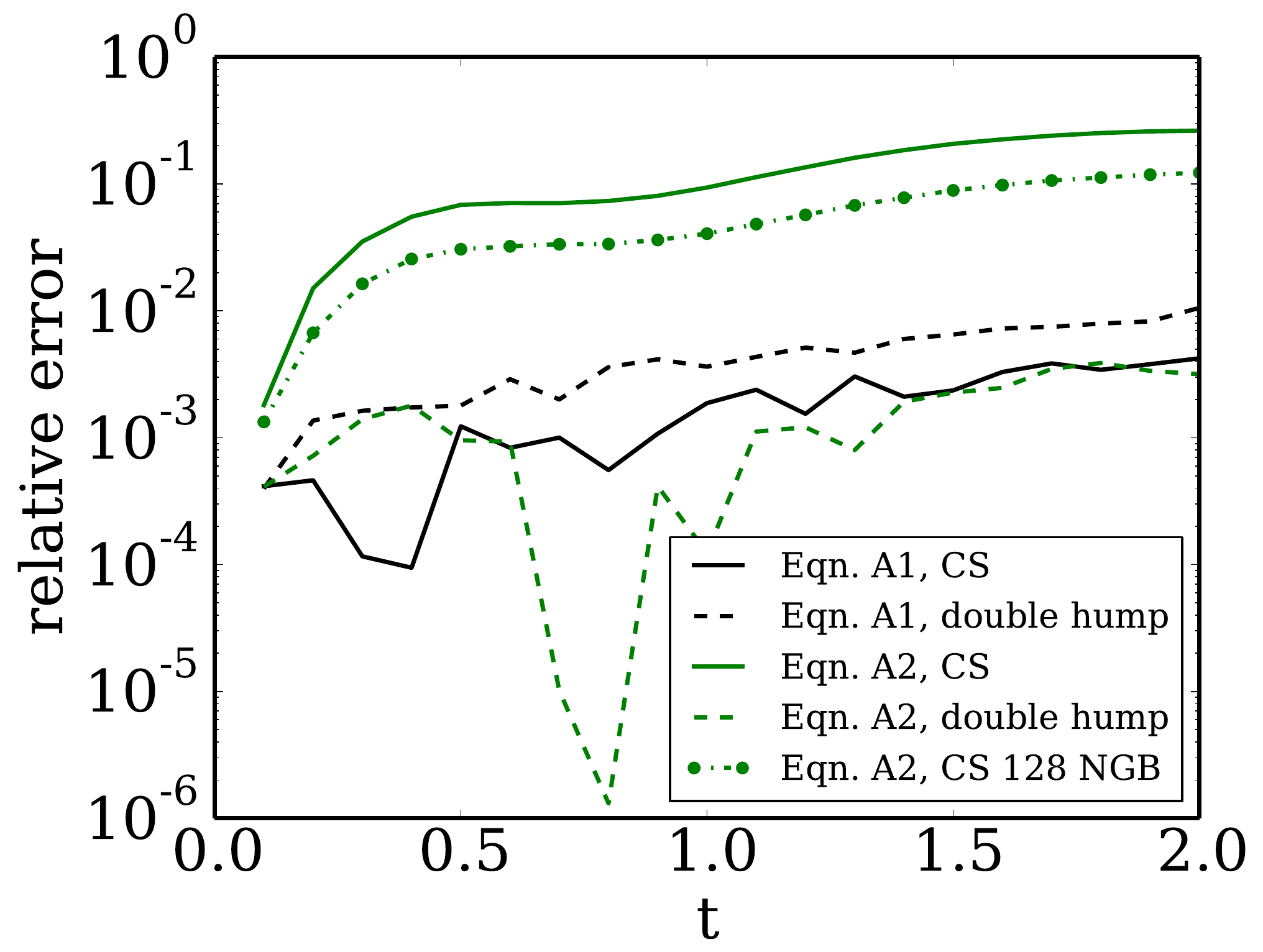}
\caption{Accuracy of different formulations of the drag law for the \textsc{dustybox} test, for the formulations given by equations~\ref{Eqn:LinearDrag} and \ref{Eqn:CentralDrag}. Both formulations produce similar results as long as appropriate kernels are used. While increasing the number of particles does not improve the accuracy of the drag forces, increasing the number of neighbours does, since this improves the local estimate of the gas properties.}
\label{Fig:DragLaws}
\end{figure}

The evaluation of drag forces in SPH requires interpolation of the gas properties to the location of the dust particles and vice versa when the feedback on the dust is important. The most straight-forward way that this can be done is via SPH interpolation at the location of the dust particles, giving a force
\begin{equation}
\vec{F}_d =  - \sum_i \frac{m_i}{\rho_i} K_{id} (\vec{v}_d - \vec{v}_i) W(r_{id}, h_d), \label{Eqn:LinearDrag}
\end{equation}
which is the approach taken by \citet{Rice2004} and used throughout this paper. However, if this approach is used for the velocity difference then the expression does not conserve angular momentum explicitly \citep{Monaghan1995,Laibe2012a}. To solve this they propose projecting the force been pairs of particles along the line between them, which gives
\begin{equation}
\vec{F}_d =  -  \nu \sum_i \frac{m_i}{\rho_i} K_{id} (\vec{v}_{di} \cdot \hat{\vec{r}}_{id}) \hat{\vec{r}}_{id} W(r_{id}, h_d), \label{Eqn:CentralDrag}
\end{equation}
where $\nu$ is the number of dimensions and $\vec{v}_{di} = \vec{v}_d - \vec{v}_i$. Equation \ref{Eqn:CentralDrag} approximates the force as the average of a series of forces acting in different directions. This means that even if the velocity field is uniform, the distribution of particles affects the force. Additionally, even if the particle distribution is regular a large number of neighbours is needed to evaluate the force accurately. \citet{Laibe2012a} suggest resolving this problem by using a double hump kernel that has the form $W(q) \propto q^2 G(q)$, where $G(q)$ approximates a Gaussian. This essentially removes the contribution of particles closest to the dust particle. Since it is these particles that drive the bias when a centrally peaked kernel is used, down weighting them is effective in removing the bias. For the original form (equation \ref{Eqn:LinearDrag}) the bias is not present and a kernel that approximates a Gaussian should be used. 

We illustrate these properties in Fig.~\ref{Fig:DragLaws}, which shows the velocity error from both schemes using either cubic spline or double hump cubic spline kernels. The performance of both schemes is similar, and as expected equation \ref{Eqn:LinearDrag} is less sensitive to the choice of kernel, although it is still more accurate for the standard cubic-spline kernel. Since we have investigated the drag forces in the test-particle limit, in which exact angular momentum conservation is not an issue, the comparable accuracy justifies our choice of force law, which was essentially motivated by facilitating a direct comparison to earlier work. However, we suggest equation \ref{Eqn:CentralDrag} should be used when the force from the dust on the gas cannot be neglected.

\label{lastpage}

\end{document}